\documentclass[journal]{IEEEtran} 
\usepackage{acronym}
\usepackage{booktabs}
\usepackage{graphicx}
\usepackage{xurl}
\usepackage[table,svgnames]{xcolor}
\usepackage{colortbl}
\usepackage{tikz}
\usepackage{orcidlink}
\usepackage{listings}
\usepackage{xcolor}

\colorlet{punct}{red!60!black}
\definecolor{background}{HTML}{EEEEEE}
\definecolor{delim}{RGB}{20,105,176}
\colorlet{numb}{magenta!60!black}

\lstdefinelanguage{json}{
    basicstyle=\normalfont\ttfamily\footnotesize,
    numbers=left, 
    showstringspaces=false,
    breaklines=true,
    frame=l, 
    literate=
     *{0}{{{\color{numb}0}}}{1}
      {1}{{{\color{numb}1}}}{1}
      {2}{{{\color{numb}2}}}{1}
      {3}{{{\color{numb}3}}}{1}
      {4}{{{\color{numb}4}}}{1}
      {5}{{{\color{numb}5}}}{1}
      {6}{{{\color{numb}6}}}{1}
      {7}{{{\color{numb}7}}}{1}
      {8}{{{\color{numb}8}}}{1}
      {9}{{{\color{numb}9}}}{1}
      {:}{{{\color{punct}{:}}}}{1}
      {,}{{{\color{punct}{,}}}}{1}
      {\{}{{{\color{delim}{\{}}}}{1}
      {\}}{{{\color{delim}{\}}}}}{1}
      {[}{{{\color{delim}{[}}}}{1}
      {]}{{{\color{delim}{]}}}}{1},
}

\definecolor{LightBlue}{HTML}{CFE2F3}
\definecolor{DarkBlue}{HTML}{6FA8DC}
\usepackage{tikz}
\newcommand{\EmptyCircle}{%
  \tikz[baseline=-0.6ex]\draw[line width=0.4pt] (0,0) circle (0.8ex);%
}
\newcommand{\HalfCircle}{%
  \tikz[baseline=-0.6ex]{
    \draw[line width=0.4pt] (0,0) circle (0.8ex);
    \fill (0,0) -- (0:0.8ex) arc (0:180:0.8ex) -- cycle;
  }%
}
\newcommand{\FullCircle}{%
  \tikz[baseline=-0.6ex]\fill (0,0) circle (0.8ex);%
}

\ifCLASSINFOpdf
\else
\fi
\usepackage{amsmath}
\usepackage{url}

\title{ORION: Intent-Aware Orchestration in Open RAN for SLA-Driven Network Management}

\author{Gabriela~da~Silva~Machado,
Gustavo~Z.~Bruno\orcidlink{0000-0002-1424-3404}, 
Alexandre~Huff\orcidlink{0000-0003-0371-4837}, 
José~Marcos~Camara~Brito\orcidlink{0000-0003-3455-8363}, and
Cristiano~B.~Both\orcidlink{0000-0002-9776-4888}

\IEEEcompsocitemizethanks{
\IEEEcompsocthanksitem Manuscript received March XX, 2026.
\IEEEcompsocthanksitem Gustavo Zanatta Bruno and José Marcos Camara Brito are with the National Telecommunications Institute (INATEL). E-mail: gustavo.zanatta@posdoc.inatel.br, brito@inatel.br.
\IEEEcompsocthanksitem Gabriela da Silva Machado and Cristiano B. Both are with the University of Vale do Rio dos Sinos (UNISINOS). E-mails: gabrielaMachado3@edu.unisinos.br, cbboth@unisinos.br.
\IEEEcompsocthanksitem Alexandre Huff is with the Federal Technological University of Paraná (UFTPR). E-mail: alexandrehuff@utfpr.edu.br.
}}

\begin{document}
\acrodef{3GPP}{3rd Generation Partnership Project}
\acrodef{5G}{Fifth Generation}
\acrodef{5GC}{5G Core}
\acrodef{A1}{A1 Interface}
\acrodef{A1AP}{A1 Interface Application Protocol}
\acrodef{ACL}{Access Control List}
\acrodef{AI}{Artificial Intelligence}
\acrodef{NLP}{Natural Language Processing}
\acrodef{NER}{Named Entity Recognition}
\acrodef{DRL}{Deep Reinforcement Learning}
\acrodef{ETSI}{European Telecommunications Standards Institute}
\acrodef{API}{Application Programming Interface}
\acrodef{CU}{Central Unit}
\acrodef{DU}{Distributed Unit}
\acrodef{E2}{E2 Interface}
\acrodef{E2Sim}{E2 Simulator}
\acrodef{FCAPS}{Fault, Configuration, Accounting, Performance, Security}
\acrodef{gNB}{gNodeB}
\acrodef{HTTP}{Hypertext Transfer Protocol}
\acrodef{IBN}{Intent-Based Network}
\acrodef{IDN}{Intent-Driven Network}
\acrodef{IETF}{Internet Engineering Task Force}
\acrodef{JSON}{JavaScript Object Notation}
\acrodef{KPI}{Key Performance Indicator}
\acrodef{ML}{Machine Learning}
\acrodef{NG-RAN}{Next Generation RAN}
\acrodef{Near-RT}{Near-Real-Time}
\acrodef{O-RAN}{Open Radio Access Network}
\acrodef{O1}{O1 Interface}
\acrodef{O2}{O2 Interface}
\acrodef{OAM}{Operations, Administration, and Maintenance}
\acrodef{ONF}{Open Networking Foundation}
\acrodef{ONOS}{Open Network Operating System}
\acrodef{PRB}{Physical Resource Block}
\acrodef{Non-RT}{Non-Real-Time}
\acrodef{QoE}{Quality of Experience}
\acrodef{QoS}{Quality of Service}
\acrodef{RAN}{Radio Access Network}
\acrodef{RFC}{Request for Comments}
\acrodef{RIC}{RAN Intelligent Controller}
\acrodef{RMIO}{RAN Management Intent Owner}
\acrodef{RMIH}{RAN Management Intent Handler}
\acrodef{RU}{Radio Unit}
\acrodef{SMO}{Service Management and Orchestration}

\acrodef{SLA}{Service Level Agreement}

\acrodef{IoT}{Internet of Things}
\acrodef{MCP}{Model Context Protocol}
\acrodef{SLO}{Service Level Objective}

\acrodef{TMF}{Telecommunications Management Forum}
\acrodef{ZSM}{Zero touch network and Service Management}
\acrodef{SDO}{Standards Development Organization}
\acrodef{RL}{Reinforcement Learning}
\acrodef{LLM}{Large Language Model}
\acrodef{AI-RAN}{AI-native RAN}
\acrodef{AI-for-RAN}{AI for RAN}
\acrodef{AI-and-RAN}{AI and RAN}
\acrodef{SON}{Self-Organizing Network}
\acrodef{UE}{User Equipment}
\acrodef{UE}{User Equipment}
\acrodef{URLLC}{Ultra-Reliable Low-Latency Communications}

\acrodef{CLA}{Closed-Loop Assurance}
\acrodef{SLO}{Service Level Objective}
\acrodef{KPM}{Key Performance Metric}
\acrodef{EI}{Enrichment Information}
\acrodef{WG}{Working Group}
\acrodef{WG2}{Working Group 2}
\acrodef{RMR}{RIC Message Router}
\acrodef{SCTP}{Stream Control Transmission Protocol}
\acrodef{E2AP}{E2 Application Protocol}
\acrodef{E2SM}{E2 Service Model}
\acrodef{PDU}{Protocol Data Unit}
\acrodef{E2Mgr}{E2 Manager}
\acrodef{OSC}{O-RAN Software Community}
\acrodef{REST}{Representational State Transfer}
\acrodef{CAMARA}{Common API Mobile And Routing Access}

\acrodef{eMBB}{Enhanced Mobile Broadband}
\acrodef{mMTC}{Massive Machine Type Communications}

\acrodef{SSE}{Server-Sent Events}
\acrodef{MIMO}{Multiple Input Multiple Output}
\acrodef{ASN.1}{Abstract Syntax Notation One}
\acrodef{PER}{Packed Encoding Rules}

\acrodef{FEACI}{Format, Explainability, Accuracy, Cost, and Inference time}

%
%

\markboth{DRAFT - DO NOT DISTRIBUTE}%
{ORION Intent-Aware Orchestration}
%



\maketitle
\bstctlcite{BSTcontrol}

\begin{abstract}
The disaggregation of the Radio Access Network (RAN) introduces architectural flexibility but significant operational complexity, necessitating automated management frameworks.
For instance, current Open RAN (O-RAN) orchestration relies on fragmented manual policies, lacking end-to-end intent assurance from high-level requirements to low-level configurations.
In this article, we propose ORION, an intent orchestration framework that integrates Large Language Models (LLMs) through the Model Context Protocol (MCP).
It translates natural language slice requests into network policies aligned with O-RAN interfaces, supporting the AI-RAN agentic orchestration and AI-for-RAN paradigms.
ORION combines an MCP-based Service Management and Orchestration (SMO) layer with a Non-Real-Time RIC rApp and a Near-Real-Time RIC xApp to enforce Physical Resource Block (PRB) quotas over E2.
We evaluate eight LLM configurations across proprietary and open-weight models on a 100-intent workload.
Claude Opus 4.5 and GPT-5 achieve 100\% policy-creation success with 99\% and 96\% slice-type classification accuracy, while GPT-5 Nano achieves 97\% policy-creation success.
The evaluation also assesses MCP tool-use quality, cost-performance trade-offs, resource consumption, and end-to-end control-path latency.
\end{abstract}


%
\IEEEpeerreviewmaketitle

%
%
\section{Introduction}\label{sec:introduction}
The \ac{5G} mobile communications ecosystem has embraced disaggregation and openness to unlock flexible service delivery~\cite{Polese2023Understanding}.
However, these same architectural properties significantly amplify the operational burden on \ac{RAN} operators.
The \ac{O-RAN} architecture formalizes this transition by splitting control across the \ac{SMO} framework, the \ac{Non-RT} \ac{RIC}, and the \ac{Near-RT} \ac{RIC}, each exposing distinct programmatic interfaces.
In practice, operators must coordinate a growing catalog of rApps and xApps across disparate control timescales ranging from non-real-time ($>1\,\mathrm{s}$) to near-real-time ($10\,\mathrm{ms}\text{--}1\,\mathrm{s}$).
Reconciling this multi-tier control hierarchy with strict \acp{SLA} and \acp{SLO} makes manual parameter provisioning and multi-interface scripting progressively intractable.
To mitigate this complexity, operators are transitioning toward \ac{AI}-driven automation and \ac{AI-RAN} architectures capable of maintaining end-to-end observability and intelligent control across disaggregated components~\cite{Habibi2024Towards,Donatti2024Survey}.

\ac{IBN} addresses \ac{O-RAN} operational complexity by translating declarative operator goals into enforceable policies~\cite{Velasco2021End,Clemm2022RFC9315}.
At the \ac{SMO}, this abstraction replaces manual formulation of vendor-specific \ac{A1} and \ac{E2} payloads.
Intent management and closed-loop frameworks from \ac{3GPP}, \ac{ETSI}, and the \ac{O-RAN} Alliance support this coupling~\cite{3GPP2025Intent,ETSI2024ZSM016,Banerjee2021Intent}.
Academic research and industry proposals increasingly extend \ac{IBN} with data-driven and \ac{LLM}-based assistants to interpret complex natural language goals~\cite{Tran2024INA,Parashar2025Intent}.
However, existing \ac{IBN} frameworks focus primarily on cloud-native core abstractions or isolated intent translations detached from \ac{O-RAN}'s multi-tier controller pipeline~\cite{Habib2023Intent,Dinh2025Towards,Wu2025LLMxApp}.
Direct translation into proprietary controller parameters risks hallucinated values and architectural coupling.
The Linux Foundation and GSMA CAMARA project supplies a standardized slice-request model as an intermediate representation between operator goals and \ac{RAN} enforcement~\cite{CAMARA2024Slice}.
Consequently, prior efforts offer limited guarantees on \ac{SLA} preservation and lack an integrated lifecycle that spans intent ingestion, schema-grounded validation, deployment, and closed-loop adaptation.
These limitations motivate our investigation of intent-aware orchestration tailored specifically to the \ac{O-RAN} control architecture.
In particular, we examine how to couple \ac{LLM}-assisted intent ingestion with \ac{O-RAN}-compliant control loops so that operator intents become deployable, monitorable policies without compromising latency, throughput, or reliability commitments.

Claude Opus 4.5 and GPT-5 achieved 100\% policy-creation success, with slice-type classification accuracies of 99\% and 96\%, respectively.
GPT-5 Nano achieved 97\% policy success at 0.49\,cents per intent, while \ac{Near-RT} \ac{RIC}/\ac{E2} control processing remained in the millisecond range.
Building upon these findings, this article presents the following contributions:

\begin{itemize}
    \item We introduce ORION, integrating \ac{AI-RAN} agentic orchestration and \ac{AI-for-RAN} across the \ac{SMO} and both \ac{RIC} tiers.
    Its \ac{CAMARA} \emph{NetworkSliceBooking} schema separates intent parsing from controllers, validates requests, and supports vendor interoperability~\cite{CAMARA2024Slice}.
    \item We design validation and composition flows across \ac{SMO} applications, rApp, and xApp, linking schema checks and policy generation to \ac{A1}/\ac{E2} control.
    These flows provide the architectural foundation for \ac{CLA} loops.
    \item We develop a prototype that combines \ac{MCP} services with \ac{A1} and \ac{E2} interfaces to translate intents into \ac{PRB} quota allocations.
    \item We introduce prompt-driven intent generation for arbitrary dataset sizes.
    Our benchmark includes 100 natural-language slice intents with reference \ac{eMBB}, \ac{URLLC}, and \ac{mMTC} labels.
    \item We compare policy creation, tool-use quality, and cost-performance trade-offs across eight \ac{LLM} configurations, including two open-weight models.
    We also measure resource overhead and end-to-end latency for edge deployment.
\end{itemize}

Sections~\ref{sec:background} and~\ref{sec:related_work} cover the background and related work.
Section~\ref{sec:architecture} presents ORION, and Section~\ref{sec:implementation} describes the prototype.
Section~\ref{sec:evaluation} evaluates the framework, and Section~\ref{sec:conclusion} concludes with future directions.

%
%
%
\section{Background}\label{sec:background}

This section establishes the foundational concepts required to understand the proposed intent-driven architecture.
We first review the \ac{O-RAN} control loops and interfaces, followed by the evolution to \ac{AI-RAN} and the standardization of Northbound telecommunication \acp{API} via \ac{CAMARA}.
Then, we examine the distinction between declarative intents and imperative policies, a consolidated intent lifecycle, and the management roles responsible for its execution.
Finally, we introduce \ac{MCP}, which underpins the \ac{LLM}-based translation engine.

\subsection{O-RAN Architectural Elements}\label{sec:bg-oran-elements}
\ac{O-RAN} disaggregates the \ac{3GPP} \ac{gNB} into virtualized \ac{CU} and \ac{DU} plus \ac{RU}~\cite{Polese2023Understanding}.
Two intelligent controllers, the \ac{Non-RT} \ac{RIC} (within the \ac{SMO}) and the \ac{Near-RT} \ac{RIC} (edge), enable policy- and \ac{ML}-driven optimization at complementary time scales (\ac{Non-RT} operating at $>1\,\mathrm{s}$, and \ac{Near-RT} at $10\,\mathrm{ms}$--$1\,\mathrm{s}$)~\cite{Niknam2022Intelligent}.
The \ac{SMO} hosts modular management functions and rApps (service-level analytics and optimization), while the \ac{Near-RT} \ac{RIC} hosts xApps that execute near-real-time control loops over radio resources.

Data and control traverse three distinct open interfaces.
First, \ac{A1} conveys guidance from the \ac{Non-RT} \ac{RIC} to the \ac{Near-RT} \ac{RIC}.
Second, \ac{E2} streams telemetry upward and distributes fine-grained control actions downward to \ac{CU}/\ac{DU}.
Finally, \ac{O1} supports \ac{FCAPS} management between the \ac{SMO} and \ac{RAN} elements~\cite{Polese2023Understanding}.
This layering splits long-horizon optimization in the \ac{Non-RT} domain from sub-second adaptation in the \ac{Near-RT} edge domain.
In this context, the openness that enables innovation also creates three coordination challenges.
First, heterogeneous vendor data models must be normalized.
Second, timing alignment between \ac{A1} and \ac{E2} control loops must avoid oscillations.
Third, intent translation must ensure consistency when multiple rApps and xApps manipulate overlapping \acp{KPI}~\cite{Habib2023Intent}.
These issues motivate the structured abstractions described next.

\subsection{Evolution to AI-RAN}\label{sec:bg-evolution-airan}
The complexity of managing these disaggregated, multi-vendor \ac{O-RAN} environments has outpaced the capabilities of traditional \ac{SON} (Self-Organizing Networks), which rely on rigid, rule-based algorithms for narrow optimization tasks. This fundamental limitation has led the industry toward the \ac{AI-RAN} paradigm, which integrates \ac{AI} natively into the network architecture rather than treating it as an overlay.

The \ac{AI-RAN} Alliance white paper defines \emph{AI-for-RAN}, \emph{AI-and-RAN}, and \emph{AI-on-RAN} in Section~3.1, with their working groups in Sections~5.1--5.3~\cite{AIRAN2024}.
These address radio optimization, shared \ac{AI}/\ac{RAN} computing, and network-hosted \ac{AI} services, respectively.
The \emph{Agentic AI Task Group} addresses autonomous network operations with \ac{AI} agents~\cite{AIRAN2025Agentic}.
ORION's \ac{SMO} intent translation aligns with this group, while its proposed \ac{Near-RT} \ac{RIC} resource control supports \emph{AI-for-RAN} objectives~\cite{AIRAN2026WG1}.

\subsection{CAMARA and Northbound API Standardization}\label{sec:bg-camara-apis}
The Linux Foundation and GSMA CAMARA project exposes vendor-neutral Northbound telecommunication \acp{API}~\cite{CAMARA2024Slice}.
ORION uses the \emph{NetworkSliceBooking} OpenAPI~3.0.3 \texttt{/sessions} resource to structure slice bookings with throughput and delay budgets, service area, availability, and duration.
This typed intermediate representation separates operator goals from controller implementations and supports argument validation before policy generation.

\subsection{Policies vs. Intents}\label{sec:bg-policies-vs-intents}
In \ac{O-RAN}, policies are structured rules: when events and conditions hold, specified actions execute to steer performance (e.g., load balancing or \ac{QoS} objectives) across scoped entities (cells, slices, \ac{UE} groups) via \ac{A1}-defined \ac{JSON} policy types \cite{Polese2023Understanding}. These policies express ``if/when $\rightarrow$ do'' logic without explicitly declaring end-state semantics. Intents, by contrast, declare desired outcomes or operational goals (the \emph{what}), omitting the procedural \emph{how} \cite{3GPP2025Intent}. A single intent may decompose into multiple policies across domains (optimization, assurance, resource allocation) \cite{Velasco2021End}.

Abstraction boundary: policies bind directly to enforcement interfaces (\ac{A1} toward the \ac{Near-RT} \ac{RIC}, \ac{O1} for management operations, and, indirectly, \ac{E2} control via xApps) \cite{Polese2023Understanding}. In contrast, intents act as a contract specifying \acp{KPI} (e.g., ``average cell-edge throughput $\geq$ $x$ Mbps within latency constraint $y$'') \cite{3GPP2025Intent}. Translation engines map such \ac{KPI}-oriented statements into one or more concrete policy objects plus monitoring queries. Policies are evaluated reactively in response to triggers, while Intentions are continuously ensured, with fulfillment engines monitoring \acp{KPI} against target states and adapting the underlying policies \cite{Banerjee2021Intent}.

Challenges include conflict detection (e.g., two intents driving contradictory \ac{PRB} allocation goals), measurability (ensuring each intent has observable success metrics), and drift handling (environmental changes invalidating earlier translations) \cite{Velasco2021End}. Therefore, effective frameworks (i) capture goals and constraints, (ii) validate feasibility against current topology and resources, (iii) translate into minimal, consistent policy sets, and (iv) maintain \ac{CLA} adjustments to preserve declared outcomes \cite{3GPP2025Intent,Banerjee2021Intent}.

\subsection{Intent Lifecycle}\label{sec:bg-intent-lifecycle}
This subsection compiles a consolidated intent lifecycle from complementary standards and specifications.
It combines fulfillment and assurance concepts from \ac{IETF} \ac{RFC}~9315 Section~6 with intent operations and reporting from \ac{3GPP} TS~28.312~\cite{Clemm2022RFC9315,3GPP2025Intent}.
ETSI GS~ZSM~016 provides closed-loop context, while TM Forum provides intent management \acp{API} and concepts~\cite{ETSI2024ZSM016,TMF2024API,TMF2022Intent}.
Together, these specifications inform a unified reference workflow rather than a verbatim six-state machine from a single standard.
Its phases are Create (register constraints), Activate (deploy policies), Monitor/Assure (check \acp{KPI}), Modify (adapt or roll back), Suspend (pause execution), and Terminate (release resources).
Transitions coordinate orchestration and support auditing through emitted events.

\begin{figure}[h]
  \centering
  \includegraphics[width=\linewidth]{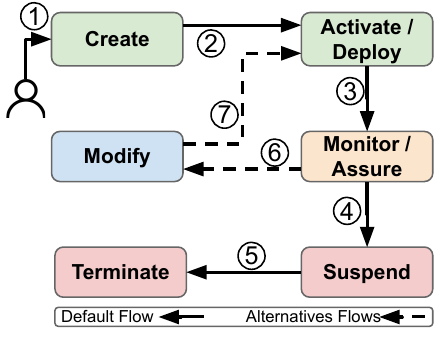}
  \caption{Consolidated intent lifecycle compiled from IETF RFC~9315~\cite{Clemm2022RFC9315}, 3GPP TS~28.312~\cite{3GPP2025Intent}, ETSI GS~ZSM~016~\cite{ETSI2024ZSM016}, and TM Forum TMF921~\cite{TMF2024API}.
  Solid arrows show forward flows (1--5), and dashed arrows show assurance feedback (6--7).}
  \label{fig:intent-lifecycle}
\end{figure}

Figure~\ref{fig:intent-lifecycle} illustrates these workflow phases and transition flows.
Solid arrows (1--5) depict the default forward progression from creation through activation, assurance, suspension, and termination.
Dashed arrows capture the \ac{CLA} feedback loop for \ac{O-RAN}.
Arrow~6 represents corrective adaptations triggered when monitored \acp{KPI} deviate from target thresholds, while arrow~7 represents the deployment or re-activation of updated policies.
\ac{CLA} evaluates whether the observed state matches the declared outcome.
Observed deviations trigger re-translation through updated \ac{A1} policies or operator notification.
This adaptive cycle maintains alignment with declared goals under dynamic network conditions.

\subsection{Roles in Intent-Driven RAN}\label{sec:bg-roles-intent-ran}
Within \ac{SMO}-based handling, the \ac{RMIO} formulates intents (goals, constraints, \ac{KPI} targets, reporting cadence), owns lifecycle authority (modify, suspend, terminate), and defines assurance requirements. The \ac{RMIH} advertises supported intent classes, interprets \ac{RMIO} submissions, decomposes them into actionable policies/configurations (\ac{A1} guidance, \ac{O1} management actions), orchestrates deployment, resolves alternative fulfillment strategies (optionally consulting the \ac{RMIO}), and performs continuous assurance.

Operationally, the \ac{RMIO} queries capability advertisements, submits a validated intent, and receives an identifier with a reporting schedule. Subsequently, the \ac{RMIO} is asynchronously notified of assurance status or required decisions (e.g., selecting among alternative resource trade-offs). The \ac{RMIH} maintains a dependency graph that maps intents to generated policies and the underlying xApp and rApp actions, enabling targeted updates rather than wholesale recomputation when only one \ac{KPI} threshold changes.

The \ac{RMIH} manages conflicts across coexisting intents (priority-based arbitration or multi-objective optimization) \cite{Velasco2021End}, adapts policies upon \ac{KPI} drift, and issues status/fulfillment reports \cite{Banerjee2021Intent}. This separation clarifies accountability (ownership vs. operational realization) and supports scalable multi-tenant automation in \ac{O-RAN}, while providing clear hooks for auditing, rollback, and intent evolution.

\subsection{Model Context Protocol}\label{sec:bg-mcp}

To realize the \ac{LLM}-driven orchestration and intent translation described previously, the intelligence engine requires a standardized mechanism to interact with the underlying \ac{O-RAN} components, such as performance monitoring logs and policy enforcement interfaces.
In this context, the \ac{MCP} represents a promising mechanism, offering an open standard protocol to connect \ac{AI} assistants with external data sources and tools, enabling the generation of context-aware responses~\cite{MCPNews}.
\ac{MCP} follows a client-server architecture in which a host (e.g., an \ac{LLM} application) establishes a connection to a server that provides access to external systems, databases, and specialized workflows~\cite{MCPdocs}.

The architecture comprises an \ac{MCP} Host, \ac{MCP} Client, and \ac{MCP} Server.
The Host manages the Client and provides the execution environment.
The Client acts as an intermediary, initiating requests to the Server and retrieving descriptions of available tools and capabilities.
The Server executes operations on external systems.
When a Host needs to perform an action, it queries the Server via the Client to discover tools, selects the appropriate one, and requests execution.
The Server performs the operation and returns the result~\cite{hou2025modelcontextprotocolmcp}.

Within the intent-driven \ac{RAN} context, \ac{MCP} bridges the cognitive capabilities of \acp{LLM} with the operational realities of the network, enabling the system to expose \ac{RMIH} functions, network controllers, and performance databases as accessible tools to the \ac{AI} agent.
Compared with provider-specific direct tool calling, \ac{MCP} standardizes discovery and invocation through an open client-server protocol over Server-Sent Events (SSE) or stdio~\cite{hou2025modelcontextprotocolmcp}.
Model providers can change without redesigning backend telecommunication services.
Operators can expose authorized tools to third-party clients across administrative domains, enabling federation scenarios.
ORION keeps backend credentials, authorization scopes, and internal routing at the server and validates arguments against CAMARA schemas before dispatch.
This boundary rejects malformed requests and limits access to private controller endpoints.
Despite architectural openness, translating high-level intents consistently and explainably into coherent multi-interface policies (\ac{A1}, \ac{O1}, \ac{E2}) while ensuring outcomes under dynamic traffic remains non-trivial.
These characteristics motivate the related work and the proposed architecture that follows.

\section{Related Work}\label{sec:related_work}

\begin{table*}[ht]
\centering
\caption{Comparative analysis of intent-driven network management frameworks across network domains: representation, translation, lifecycle/assurance, slicing, O-RAN, and AI-RAN}
\label{tab:intent-related}
\renewcommand{\arraystretch}{1.15}
\resizebox{\textwidth}{!}{%
\begin{tabular}{c c ll cccccc}
\hline
\cellcolor{DarkBlue!80}\textbf{Year} &
\cellcolor{DarkBlue!80}\textbf{Domain} &
\multicolumn{2}{c}{\cellcolor{DarkBlue!80}\textbf{Paper}} &
\cellcolor{DarkBlue!80}\textbf{Intent Rep.} &
\cellcolor{DarkBlue!80}\textbf{Translation} &
\cellcolor{DarkBlue!80}\textbf{Lifecycle/Assur.} &
\cellcolor{DarkBlue!80}\textbf{Slicing} &
\cellcolor{DarkBlue!80}\textbf{O-RAN} &
\cellcolor{DarkBlue!80}\textbf{AI-RAN} \\
\hline
\rowcolor{white}
\textcolor{gray}{2020} & RAN & \cite{Abbas2020IBN} & Graph-based RAN slice orchestration & \HalfCircle & \FullCircle & \EmptyCircle & \FullCircle & \EmptyCircle & \EmptyCircle \\
\rowcolor{LightBlue!100}
\textcolor{gray}{2021} & RAN & \cite{Banerjee2021Intent} & Rule-pipeline intent to KPI commands & \HalfCircle & \FullCircle & \EmptyCircle & \EmptyCircle & \EmptyCircle & \EmptyCircle \\
\rowcolor{white}
\textcolor{gray}{2021} & E2E & \cite{Velasco2021End} & End-to-end IBN (standardized model) & \HalfCircle & \HalfCircle & \EmptyCircle & \EmptyCircle & \HalfCircle & \EmptyCircle \\
\rowcolor{LightBlue!100}
\textcolor{gray}{2021} & Smart Grid & \cite{Mehmood2021Smart} & Intent-driven slice mgmt (power grid) & \FullCircle & \HalfCircle & \HalfCircle & \FullCircle & \EmptyCircle & \EmptyCircle \\
\rowcolor{white}
\textcolor{gray}{2022} & E2E & \cite{Xie2022Multi} & TMF-aligned intent for E2E slicing & \FullCircle & \HalfCircle & \HalfCircle & \FullCircle & \HalfCircle & \EmptyCircle \\
\rowcolor{LightBlue!100}
\textcolor{gray}{2023} & Private 5G & \cite{Mcnamara2023NLP} & NLP-powered private 5G management & \FullCircle & \HalfCircle & \EmptyCircle & \HalfCircle & \EmptyCircle & \EmptyCircle \\
\rowcolor{white}
\textcolor{gray}{2023} & O-RAN & \cite{Habib2023Intent} & Hierarchical DRL for O-RAN xApp orchestration & \HalfCircle & \HalfCircle & \EmptyCircle & \HalfCircle & \FullCircle & \EmptyCircle \\
\rowcolor{LightBlue!100}
\textcolor{gray}{2023} & RAN & \cite{Zhang2023MultiAgent} & Intent-driven RAN slice orchestration (MARL) & \HalfCircle & \HalfCircle & \EmptyCircle & \FullCircle & \EmptyCircle & \EmptyCircle \\
\rowcolor{white}
\textcolor{gray}{2023} & 5G Slicing & \cite{Abdellatif2023Intelligent} & Intelligent-Slicing: AI-Assisted Slicing Framework & \HalfCircle & \HalfCircle & \EmptyCircle & \FullCircle & \EmptyCircle & \EmptyCircle \\
\rowcolor{LightBlue!100}
\textcolor{gray}{2023} & E2E & \cite{Gharehgoli2023AI} & AI-based Resource Allocation in E2E Slicing & \HalfCircle & \HalfCircle & \EmptyCircle & \FullCircle & \EmptyCircle & \EmptyCircle \\
\rowcolor{white}
\textcolor{gray}{2024} & RAN & \cite{Nahum2024Intent} & Intent-aware RRS for RAN slicing (RL) & \FullCircle & \HalfCircle & \EmptyCircle & \FullCircle & \EmptyCircle & \EmptyCircle \\
\rowcolor{LightBlue!100}
\textcolor{gray}{2024} & O-RAN & \cite{Habibi2024Towards} & AI/ML-driven SMO framework in O-RAN & \HalfCircle & \HalfCircle & \EmptyCircle & \HalfCircle & \FullCircle & \EmptyCircle \\
\rowcolor{white}
\textcolor{gray}{2024} & O-RAN & \cite{Tran2024INA} & INA-Infra: open intent-driven slicing stack & \FullCircle & \HalfCircle & \HalfCircle & \FullCircle & \FullCircle & \EmptyCircle \\
\rowcolor{LightBlue!100}
\textcolor{gray}{2024} & RAN & \cite{Gopal2024AdapShare} & AdapShare: RL-based intent spectrum sharing & \HalfCircle & \FullCircle & \EmptyCircle & \HalfCircle & \FullCircle & \EmptyCircle \\
\rowcolor{white}
\textcolor{gray}{2024} & O-RAN & \cite{Bonati2024PACIFISTA} & PACIFISTA: O-RAN conflict characterization & \EmptyCircle & \HalfCircle & \HalfCircle & \EmptyCircle & \FullCircle & \EmptyCircle \\
\rowcolor{LightBlue!100}
\textcolor{gray}{2024} & Core/E2E & \cite{Huang2024ChatNet} & ChatNet: Embodied tool-use network LLM & \FullCircle & \FullCircle & \HalfCircle & \HalfCircle & \EmptyCircle & \EmptyCircle \\
\rowcolor{white}
\textcolor{gray}{2024} & 6G / O-RAN & \cite{Chatzistefanidis2024Maestro} & Maestro: Collaborative multi-plane intent automation & \FullCircle & \FullCircle & \HalfCircle & \HalfCircle & \HalfCircle & \HalfCircle \\
\rowcolor{LightBlue!100}
\textcolor{gray}{2025} & DC / SDN & \cite{Jacobs2025Trust} & Intent trust and integrity assessment in IBN & \FullCircle & \FullCircle & \HalfCircle & \EmptyCircle & \EmptyCircle & \EmptyCircle \\
\rowcolor{white}
\textcolor{gray}{2025} & DC / SDN & \cite{Hossain2025NetIntent} & NetIntent: Multi-agent SDN intent automation & \FullCircle & \FullCircle & \HalfCircle & \EmptyCircle & \EmptyCircle & \EmptyCircle \\
\rowcolor{LightBlue!100}
\textcolor{gray}{2025} & Core / NFV & \cite{Wei2025INTA} & INTA: Multi-agent translation for network config & \FullCircle & \FullCircle & \EmptyCircle & \HalfCircle & \EmptyCircle & \EmptyCircle \\
\rowcolor{white}
\textcolor{gray}{2025} & Softwarized & \cite{Angi2025LLNet} & LLNet: SLM intent for softwarized devices & \HalfCircle & \FullCircle & \EmptyCircle & \EmptyCircle & \EmptyCircle & \EmptyCircle \\
\rowcolor{LightBlue!100}
\textcolor{gray}{2025} & O-RAN & \cite{Lee2025LLMhRIC} & LLM-hRIC: LLM-guided Non-RT + RL Near-RT & \HalfCircle & \FullCircle & \EmptyCircle & \HalfCircle & \FullCircle & \HalfCircle \\
\rowcolor{white}
\textcolor{gray}{2025} & O-RAN & \cite{Wu2025LLMxApp} & LLM-xApp: NL-to-slice translation at Near-RT & \HalfCircle & \FullCircle & \EmptyCircle & \FullCircle & \FullCircle & \HalfCircle \\
\rowcolor{LightBlue!100}
\textcolor{gray}{2025} & RAN + Core & \cite{Dinh2025Towards} & FEACI: translation quality scoring (RAN+Core) & \HalfCircle & \HalfCircle & \EmptyCircle & \HalfCircle & \HalfCircle & \EmptyCircle \\
\rowcolor{white}
\textcolor{gray}{2025} & O-RAN & \cite{DelaCruz2025RSLAQ} & RSLAQ: SLA-driven QoS with A1/E2 loops & \HalfCircle & \FullCircle & \HalfCircle & \FullCircle & \FullCircle & \EmptyCircle \\
\rowcolor{LightBlue!100}
\textcolor{gray}{2025} & Cloud / Core & \cite{Parashar2025Intent} & Intent APIs for lifecycle control & \HalfCircle & \FullCircle & \HalfCircle & \EmptyCircle & \EmptyCircle & \EmptyCircle \\
\rowcolor{white}
\textcolor{gray}{2025} & AI-RAN & \cite{Chatzistefanidis2025MXAI} & MX-AI: Agentic observability and control for AI-RAN & \FullCircle & \FullCircle & \HalfCircle & \HalfCircle & \FullCircle & \FullCircle \\
\rowcolor{LightBlue!100}
\textbf{2026} & \textbf{Open RAN} & \textbf{This Work} & E2E Intent Orchestration & \FullCircle & \FullCircle & \HalfCircle & \FullCircle & \FullCircle & \FullCircle \\
\hline
\end{tabular}}
\footnotesize
\FullCircle\,Yes\enspace\HalfCircle\,Partial\enspace\EmptyCircle\,No
\end{table*}

This section surveys intent-driven network management across telecommunication domains and \ac{O-RAN}.
We first examine the evolving standards landscape that formalizes operational roles and lifecycle definitions for intent handling.
Subsequently, we review recent advances in integrating \acp{LLM} across diverse network segments, encompassing data center and software-defined networks, core and network function virtualization infrastructures, and transport networks.
We then analyze intent-driven proposals in \ac{O-RAN} across six architectural dimensions:
(i)~\textit{intent representation}, evaluating how declarative goals and constraints are expressed,
(ii)~\textit{translation and interpretation}, examining mechanisms that map high-level objectives into executable policies,
(iii)~\textit{lifecycle and assurance}, reflecting continuous monitoring, closed-loop validation, and conflict remediation,
(iv)~\textit{network slicing scope}, detailing the orchestration of dedicated virtual resources,
(v)~\textit{\ac{O-RAN}}, indicating integration with standard \ac{SMO} and \ac{RIC} control loops, and
(vi)~\textit{\ac{AI-RAN}}, capturing alignment with \ac{AI-RAN} agentic orchestration and \ac{AI-for-RAN} workflows.
Table~\ref{tab:intent-related} provides a comparative taxonomy of the surveyed literature across these dimensions, where \FullCircle~indicates comprehensive or native support, \HalfCircle~indicates partial or non-native support, and \EmptyCircle~indicates that the dimension is not addressed.

\textbf{Standards landscape.}
The treatment of intent has matured across international \acp{SDO}.
In \ac{TMF}, the TMF921 Intent Management \ac{API} (v5.0.0) binds to the \ac{TMF} Intent Ontology, establishing schemas for exchanging machine-tractable intent objects~\cite{TMF2024API,TMF2022Intent}.
In \ac{3GPP}, TS~28.312 (Rel-18) formalizes quantifiable intents, feasibility checks, fulfillment reports, and priority semantics~\cite{3GPP2025Intent,3GPP2023Web}.
\ac{ETSI} \ac{ZSM} defines the architectural reference model for intent owner and handler roles in closed-loop automation~\cite{ETSI2024ZSM}.
Furthermore, \ac{ETSI} GS~\ac{ZSM}~016 defines a three-level conflict taxonomy with concrete detect-report-judge-fulfill interaction sequences~\cite{ETSI2024ZSM016}.
From the \ac{O-RAN} perspective, \ac{ETSI} TS~103~985 specifies mechanisms for managing \ac{O-RAN} WG2 \ac{A1} policy and \ac{EI} job lifecycles~\cite{ETSI2024TS103985}.
In complementary fashion, \ac{IETF} \ac{RFC}~9315 formalizes the conceptual foundations of \ac{IBN} and decouples intent fulfillment from runtime assurance~\cite{Clemm2022RFC9315}.
Together, these specifications establish the operational contract for intent handling but stop short of an end-to-end reference implementation for disaggregated \ac{RAN} architectures.

\textbf{\ac{LLM} integration across network segments.}
In data center and software-defined networks, Jacobs et al.~\cite{Jacobs2025Trust} examine intent integrity, while Hossain and Aljoby's NetIntent~\cite{Hossain2025NetIntent} combines multi-agent control, feedback, and IBNBench evaluation.
Angi et al.'s LLNet~\cite{Angi2025LLNet} uses small language models for resource-constrained device control.
In core networks, Huang et al.'s ChatNet~\cite{Huang2024ChatNet} accesses diagnostic tools, while Wei et al.'s INTA~\cite{Wei2025INTA} translates requirements into multi-domain slicing and virtual network function configurations.
Optical and transport studies combine \acp{LLM} with digital twins for routing, path computation, and survivability checks.

\textbf{Intent representation and translation in \ac{RAN}.}
Prior radio access proposals span natural-language interfaces, structured templates, and standards-based ontologies.
Natural-language interfaces appear in several works~\cite{Mcnamara2023NLP,Mehmood2021Smart,Habib2023Intent}, often coupled to guided forms to constrain linguistic ambiguity when specifying performance goals~\cite{Habib2023Intent}.
Template- and rule-oriented formats are evaluated in recent investigations~\cite{Banerjee2021Intent,Velasco2021End}, where operators populate predefined fields that map to declarative control policies.
Adopting standards-aligned representations, Xie et al.~\cite{Xie2022Multi} and Nahum et al.~\cite{Nahum2024Intent} employ the \ac{TMF} intent model, easing programmatic decomposition into network actions.
Translation mechanisms across the literature rely on rule pipelines~\cite{Banerjee2021Intent}, graph reasoning~\cite{Abbas2020IBN,Zhang2023MultiAgent}, or external machine-learning modules~\cite{Mcnamara2023NLP}.
Abdellatif et al.~\cite{Abdellatif2023Intelligent} examine machine-learning assistance for 5G slicing, while Gharehgoli et al.~\cite{Gharehgoli2023AI} optimize slice resource allocation under stochastic demand.
Recently, \acp{LLM} have been embedded directly into \ac{O-RAN} control loops.
Wu et al. introduce \ac{LLM}-xApp to translate dynamic user \ac{QoS} requirements into slice configurations at near-real-time timescales~\cite{Wu2025LLMxApp}.
Lee et al. propose \ac{LLM}-hRIC, combining Non-RT \ac{LLM} guidance with Near-RT reinforcement learning~\cite{Lee2025LLMhRIC}.
Tran et al. demonstrate INA-Infra, an open stack linking Nephio orchestration with rApps and xApps~\cite{Tran2024INA}.
Dinh et al. evaluate translation accuracy across combined \ac{RAN} and core domains through the \ac{FEACI} benchmark~\cite{Dinh2025Towards}.

\textbf{Lifecycle, assurance, and agentic orchestration.}
Xie et al.~\cite{Xie2022Multi} and Parashar~\cite{Parashar2025Intent} analyze lifecycle procedures (create, modify, suspend, and terminate) and handler interfaces.
Mehmood et al.~\cite{Mehmood2021Smart} maintain closed-loop assurance by tracking performance metrics and adapting slice parameters.
Runtime assurance and conflict remediation in \ac{O-RAN} have also received recent attention.
DelaCruz et al. propose RSLAQ, which maps service agreements to \acp{KPI} over \ac{A1} and \ac{E2} interfaces~\cite{DelaCruz2025RSLAQ}.
Bonati et al. present PACIFISTA, profiling rApp and xApp interactions to predict resource conflicts and performance degradation~\cite{Bonati2024PACIFISTA}.
Maestro mediates conflicting stakeholder intents~\cite{Chatzistefanidis2024Maestro}.
MX-AI organizes \ac{SMO} agents for observability and Day-0/1/2+ workflows, including typed tool outputs, schemas, validation, and \ac{A1}/\ac{E2} control~\cite{Chatzistefanidis2025MXAI}.

\textbf{Architectural synthesis and positioning.}
ORION combines \ac{MCP} tool discovery and invocation, a CAMARA \textit{NetworkSliceBooking} intermediate representation, and deterministic rApp mapping to \ac{A1} policies.
Compared with Maestro's negotiation and MX-AI's observability and blueprint management, ORION evaluates translation, classification, and policy creation for slice intents~\cite{Chatzistefanidis2024Maestro,Chatzistefanidis2025MXAI}.
The prototype enforces \ac{PRB} quotas through \ac{E2SM}-RC control and measures processing latency along the \ac{SMO}-to-\ac{E2} control path.

%
%
%
\section{Proposed Architecture: ORION}\label{sec:architecture}
This section presents ORION, an intent-aware orchestration pipeline for \ac{O-RAN} that bridges natural-language intents and deployable \ac{RAN} policies.
ORION couples \ac{LLM}-assisted \ac{MCP} services with \ac{SMO}, \ac{Non-RT} \ac{RIC}, and \ac{Near-RT} \ac{RIC} functions through \ac{O-RAN} interfaces.
It reuses the CAMARA \emph{NetworkSliceBooking} \ac{API} schema as an intermediate representation for slice requests~\cite{CAMARA2024Slice}.

\subsection{Design Objectives}\label{sec:arch-objectives}
ORION pursues five core objectives.
(i) \textit{Automation with \ac{SLA}-awareness} aims to express requested latency, throughput, and reliability constraints in schema-validated specifications.
(ii) \textit{Modularity} decouples intent assistance, interpretation, composition, and policy handling, enabling each block to evolve independently.
(iii) \textit{Role alignment and standardized representation} maps \ac{SMO} and \ac{RIC} responsibilities using \ac{RMIO} and \ac{RMIH} semantics, and adopts \ac{CAMARA} open schemas to reduce vendor coupling.
(iv) \textit{\ac{O-RAN} interface alignment} integrates the \ac{Non-RT} \ac{RIC} for \ac{A1} policy distribution and the \ac{Near-RT} \ac{RIC} and xApps for \ac{E2} control.
Finally, (v) \textit{extensibility} supports prospective load-balancing or traffic-steering tools through alternative schemas and policy templates.

\subsection{High-Level Architecture Overview}\label{sec:arch-overview}
\begin{figure}[!h]
	\centering
	\includegraphics[width=0.95\linewidth]{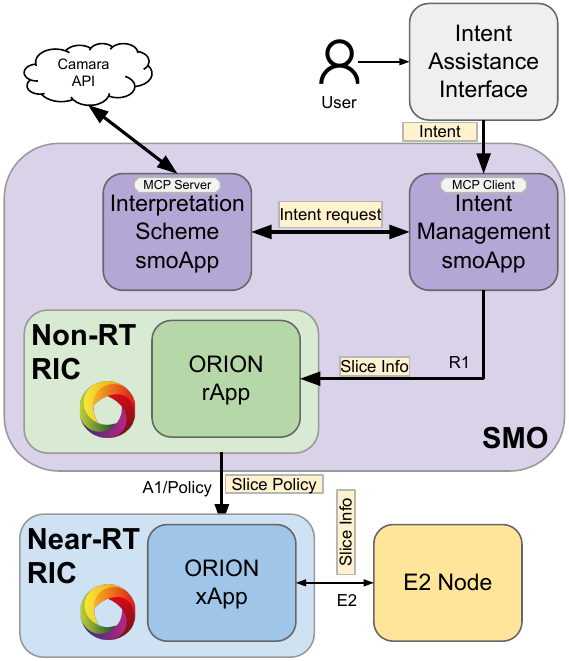}
	\caption{ORION high-level architecture spanning SMO, Non-RT RIC, and Near-RT RIC.
	The MCP Client and Server (\emph{smoApps}) structure intents using CAMARA's NetworkSliceBooking schema, while the ORION rApp composes policies and the ORION xApp enforces them via A1 and E2.}
	\label{fig:orion-architecture}
\end{figure}

Figure~\ref{fig:orion-architecture} shows ORION's blocks spanning the \ac{SMO}, \ac{Non-RT} \ac{RIC}, and \ac{Near-RT} \ac{RIC} layers.
An \ac{MCP} Client acting as a \emph{smoApp} receives operator or higher-layer intents and invokes the \ac{LLM} with tools advertised by the \ac{MCP} Server, also a \emph{smoApp}.
For slicing, the server exposes a CAMARA \emph{NetworkSliceBooking} \ac{API} consumer.
Within the \ac{SMO}, the client orchestrates tool invocation and schema validation before policy creation.

The ORION rApp maps structured requests into policy templates and \ac{SLA} objectives, such as throughput and latency budgets.
The \ac{Non-RT} \ac{RIC} publishes policies via \ac{A1}, while the ORION xApp derives \ac{PRB} quotas from cell configuration using a proportional heuristic.
The prototype applies these quotas through \ac{E2} toward an emulated \ac{gNB}.

\subsection{Component Details}\label{sec:arch-components}
The \ac{MCP} Client and Server jointly realize the intent assistance plane.
The client orchestrates the \ac{LLM} workflow and returns tool outcomes to the \ac{SMO}.
It maintains conversation state and applies prompt and schema guardrails to constrain generated tool arguments.
The server exposes tool endpoints and domain context to the \ac{LLM}.
For slicing, the server supplies a translator aligned with \ac{CAMARA}'s \textit{NetworkSliceBooking} \ac{API}\footnote{\url{https://github.com/camaraproject/NetworkSliceBooking}}~\cite{CAMARA2024Slice}.
This translator accepts an intent payload, builds a compliant \ac{JSON} POST, and returns a structured result.
CAMARA's \textit{NetworkSliceBooking} model supplies a common representation of slice \ac{QoS}, service area, and duration between language interpretation and vendor-specific controllers.
Models derived from CAMARA OpenAPI validate mandatory attributes, types, bounds, and units at the \ac{MCP} boundary before policy composition.
Under the \ac{MCP} Host's orchestration, the client invokes the selected server tool, which submits the booking request and returns diagnostics for clarification.
Compatible interfaces support prospective multi-operator integration, subject to authorization and backend validation.

The \ac{LLM} performs schema-constrained tool selection, semantic parameter slot filling, and slice-type classification in a bounded two-step workflow rather than open-ended multi-step reasoning~\cite{Qin2024ToolLLM}.
It uses the operator intent, system prompt rules, and advertised CAMARA OpenAPI~3.0.3 arguments to select a booking tool and populate its fields.
Generated \ac{JSON} arguments undergo schema validation, while omitted optional values remain null for deterministic rApp default binding.
The task extends token-level \ac{NER} through semantic extraction and unit normalization, restricting generation to tool arguments and slice-family labels rather than executable controller code.

The ORION rApp maps validated structured requests into policy templates containing requested \ac{SLA} objectives.
The architectural design includes throughput, latency, and reliability budgets, scheduler selection, \ac{QoS} classes, and monitoring metrics.
The \ac{Non-RT} \ac{RIC} distributes policies through \ac{A1}, while the ORION xApp translates policy objectives into \ac{E2} control requests.
The prototype applies \ac{PRB} quotas to emulated scheduler partitions.
In the architecture, \ac{KPM} monitoring can trigger policy recomputation or rollback through rApp and xApp coordination.

\subsection{End-to-End Intent Workflow}\label{sec:arch-workflow}
These six workflow stages organize ORION's lifecycle architecture, informed by \ac{3GPP} TS~28.312~\cite{3GPP2025Intent} and ETSI GS~ZSM~016~\cite{ETSI2024ZSM016}.
(i)~\emph{Ingestion}: an operator issues a slice intent, and the \ac{MCP} Client initializes the \ac{LLM} session.
(ii)~\emph{Translation}: the \ac{LLM} calls the \ac{MCP} Server tool, which structures the request per \ac{CAMARA} NetworkSliceBooking and returns \ac{JSON}.
(iii)~\emph{Validation and Clarification}: to resolve ambiguous or underspecified intents, the workflow combines prompt nullability rules with \ac{CAMARA} schema validation and backend admission checks.
For partially specified intents, omitted optional parameters remain null so the ORION rApp binds them deterministically to \ac{3GPP} baseline service profiles and 5QIs without model guessing.
Conversely, intents that remain ambiguous due to missing mandatory attributes, contradictory constraints, invalid bounds, or denied admission are rejected before policy composition, and the system returns structured diagnostics to the \ac{MCP} Client.
The architecture injects these diagnostics into the \ac{LLM} context to prompt the operator for clarification.
(iv)~\emph{Composition}: the ORION rApp turns the validated intent into policy templates and parameter values.
(v)~\emph{Deployment}: the \ac{Non-RT} \ac{RIC} publishes policies via \ac{A1} and the ORION xApp enforces them through \ac{E2} toward \acp{gNB}.
(vi)~\emph{Real-Time Monitoring and Closed-Loop Adaptation}: the architecture separates \ac{E2SM}-KPM telemetry to the \ac{Near-RT} \ac{RIC} from \ac{O1} performance reports to the \ac{SMO}.
Near-real-time control ($10\,\mathrm{ms}\text{--}1\,\mathrm{s}$) uses xApp \ac{PRB} updates through \ac{E2SM}-RC Control Style~2, Action ID~6.
Persistent \ac{SLA} deviations prompt \ac{A1} status feedback and rApp policy recomputation or rollback in the non-real-time loop ($>1\,\mathrm{s}$).

\subsection{Extensibility and Standards Alignment}\label{sec:arch-extensibility}
Roles map to \ac{RMIO} and \ac{RMIH}: the operator or service assumes \ac{RMIO}, while the \ac{MCP} Client and ORION rApp act as \ac{RMIH}.
ORION places intent management and composition in the \ac{SMO}/\ac{Non-RT} \ac{RIC}, with \ac{A1}/\ac{E2} control interfaces.
This placement is consistent with \ac{3GPP} TS~28.312~\cite{3GPP2025Intent}, ETSI GS~ZSM~016~\cite{ETSI2024ZSM016}, and TM Forum TMF921~\cite{TMF2024API}.
By treating \ac{CAMARA} \acp{API} as interchangeable schema backends, the same pattern generalizes to other intents (e.g., load balancing or traffic steering) by swapping the tool and policy templates without altering the core workflow.
For prospective federation, an operator could expose authorized slicing tools through an \ac{MCP} Server to third-party enterprise or partner \ac{LLM} agents acting as \ac{MCP} Clients.
Tool discovery and schema-based invocation separate external requests from internal \ac{SMO} endpoints and controller credentials.
This scenario requires cross-domain authorization and compatible backends.

In summary, the ORION architecture establishes a cohesive, intent-driven control plane that bridges natural language abstractions and low-level \ac{RAN} parameterization.
By decoupling \ac{LLM} inference in the \ac{SMO} from deterministic \ac{PRB} control in the \ac{Near-RT} \ac{RIC}, the architecture separates language interpretation from resource configuration.
Having outlined the theoretical framework, the subsequent section details the practical implementation and the proof-of-concept prototype employed to validate this end-to-end workflow.


\section{Implementation and Prototype}\label{sec:implementation}

\begin{figure*}[!h]
    \centering
    \includegraphics[width=0.9\linewidth]{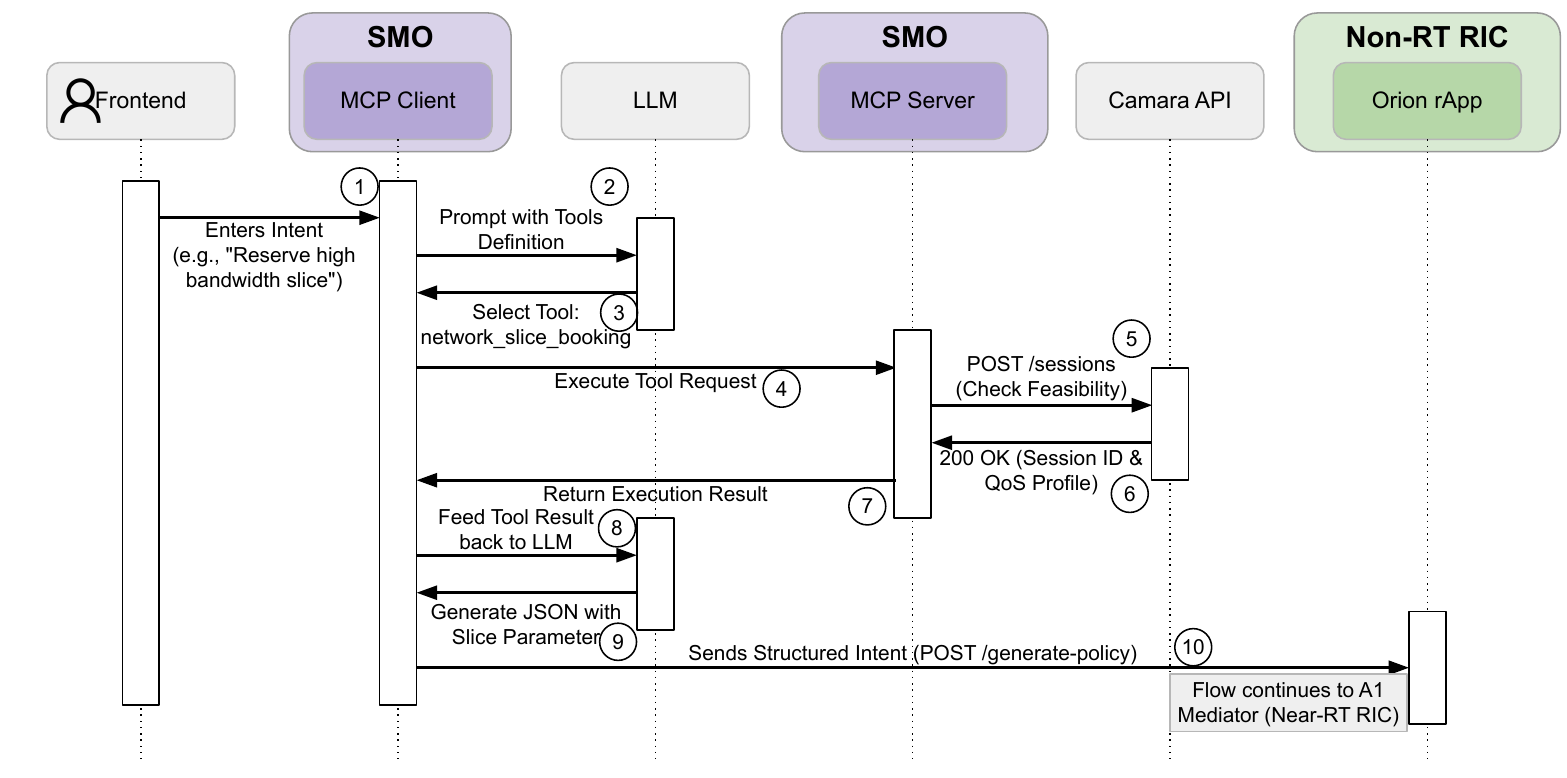}
    \caption{Non-RT RIC Intent Execution Flow.}
    \label{fig:intent_flow}
\end{figure*}

We developed a proof-of-concept prototype for the \textit{Creation} and \textit{Activation} phases of the intent-driven architecture.
The architecture in Section~\ref{sec:architecture} encompasses the full lifecycle, including proposed assurance mechanisms.
The prototype translates natural-language requests into \ac{A1} policies and applies \ac{PRB} quotas through emulated \ac{E2} control.
Containerized microservices integrate the \ac{LLM}/\ac{MCP} orchestration agent with \ac{O-RAN} control components.
The key components of the prototype are detailed as follows.

\subsection{Frontend (Streamlit)}\label{sec:impl-frontend}
A Web-based user interface built with the Streamlit framework allows users to enter natural language intents.
The interface renders the conversation history, maintaining distinct bubbles for user interaction, while persisting state across sessions.
Communication with the \ac{MCP} Client is handled via asynchronous \ac{HTTP} POST requests targeting the \texttt{/intent} endpoint.
\subsection{SMO}\label{sec:impl-smo}
The \ac{SMO} uses \ac{MCP} to coordinate slice provisioning and activation.
The \ac{MCP} Client is built with FastAPI and integrates multiple \ac{LLM} providers via a unified interface.
It manages the system prompt and injects an instruction requesting \ac{JSON} outputs.
An \ac{SSE} connection to the \ac{MCP} Server carries tool execution requests and results.
The \ac{MCP} Server is developed using the Python \texttt{mcp} library to provide a standardized interface for network functions.
It hosts the \texttt{create\_session} tool, which acts as a proxy by receiving Pydantic-validated arguments and forwarding them to the underlying Slice \ac{API}.
To handle ambiguous or underspecified requests, CAMARA-derived Pydantic models reject missing mandatory fields, invalid types, and out-of-range bounds before dispatch, while preserving omitted optional fields as null for downstream defaults.
The \ac{MCP} Server returns schema-validation and booking-backend admission failures as structured diagnostics to the client.
Prototype admission uses a CAMARA mock session quota, while \texttt{e2sim-rc} separately guards aggregate \ac{PRB} capacity.

\subsection{\texorpdfstring{\acl{LLM}}{Large Language Model}}\label{sec:impl-llm}
Within the \ac{SMO}, integrated \ac{LLM} providers (OpenAI GPT, Google Gemini, Anthropic Claude, and hosted Qwen3 models) execute two-step inference, separating argument extraction from policy computation.
First, the \ac{LLM} uses \ac{MCP} descriptors advertised over \ac{SSE} and eight field-population rules to select \texttt{create\_session} and generate arguments, leaving unmentioned fields \texttt{null}.
The system prompt prohibits fabricating default or minimum values when an intent is underspecified or ambiguous.
After tool confirmation, a second inference classifies the \ac{QoS} profile as \ac{eMBB}, \ac{URLLC}, or \ac{mMTC} for xApp scheduling.
Generation is confined to arguments and labels, while the rApp and xApp compute policies and encode \ac{E2SM}-RC messages deterministically.

\subsection{CAMARA Mock API}\label{sec:impl-camara-mock}
A mock server implementing the CAMARA \textit{Network Slice Booking} \ac{API} was developed to expose network resources.
The core endpoint, \texttt{POST /sessions}, accepts booking requests adhering to OpenAPI 3.0.3 schemas.
Admission control tracks active sessions in an in-memory dictionary.
Requests are rejected with \texttt{429 TOO\_MANY\_REQUESTS} if the active session count exceeds a predefined threshold.

\subsection{ORION rApp (Non-RT RIC)}\label{sec:impl-rapp}
The ORION rApp exposes a \texttt{POST /generate-policy} endpoint to receive structured intent data.
It maps structured requests into \ac{A1} policy objectives, such as throughput limits, delay budgets, and packet error-rate targets.
The xApp computes \ac{PRB} quotas from these objectives and the target cell configuration.
The final output is an \ac{A1} Policy \ac{JSON} payload containing the slice ID, scope, and these specific \ac{SLO} objectives.
Crucially, the policy also includes the \texttt{slice\_type}, e.g., \ac{URLLC}, \ac{eMBB}.
This semantic tag guides the xApp's scheduler selection, such as Earliest Deadline First for \ac{URLLC} and Proportional Fair for \ac{eMBB}.
Raw \ac{QoS} values alone do not fully express this service classification.
An example of such a generated policy payload is shown in Listing~\ref{lst:policy_json}.

\begin{minipage}{\columnwidth}
\begin{lstlisting}[language=json, caption={Example A1 Policy Instance generated by the ORION rApp.}, label={lst:policy_json},
    breaklines=true,
    columns=fullflexible,
    frame=single,
    linewidth=0.92\columnwidth,
    xleftmargin=0.05\columnwidth,
    framexleftmargin=1.5em,
    captionpos=b,
    resetmargins=true
    ]
{   "ric_id": "ric4",
    "policy_id": "48782",
    "service_id": "intentSlice",
    "policy_data": {
        "scope": {
            "sliceId": {
                "sst": 1,
                "sd": "456DEF",
                "plmnId": {
                    "mcc": "724",
                    "mnc": "11" },
                "nci": 1 },
            "sliceType": "mMTC" },
        "sliceSlaObjectives": {
            "maxDlThptPerUe": 50000,
            "maxUlThptPerUe": 25000,
            "maxDlThptPerSlice": 300000000,
            "maxUlThptPerSlice": 150000000 } },
    "policytype_id": 10002 }
\end{lstlisting}
\end{minipage}

\subsection{SMO and Non-RT RIC Workflow}\label{sec:impl-workflow-smo-nonrt}

Figure~\ref{fig:intent_flow} shows the interaction flow within the \ac{SMO} and \ac{Non-RT} \ac{RIC}.
The intent translation process follows a sequence initiated by the user's intent declaration.
The workflow begins when the user inputs a natural language intent via the Streamlit front end (\textbf{Step~1}).
The \ac{MCP} Client captures this input and constructs a prompt for the \ac{LLM}, including the tool definitions (\textbf{Step~2}).
Based on the prompt, the \ac{LLM} selects the \texttt{create\_session} tool and extracts explicit request arguments while keeping unmentioned optional fields \texttt{null} (\textbf{Step~3}).
Next, the \ac{MCP} Client formally requests the execution of this tool from the \ac{MCP} Server (\textbf{Step~4}).
The \ac{MCP} Server validates the arguments against the CAMARA schema and forwards valid booking requests to the CAMARA mock \ac{API} (\textbf{Step~5}), which checks its session quota.
Ambiguous requests that fail schema validation at Step~4 (e.g., missing mandatory fields or invalid bounds) or admission control at Step~5 (e.g., \texttt{429 TOO\_MANY\_REQUESTS}) return structured diagnostics through the \ac{MCP} Server.
The architecture uses this feedback for \ac{LLM}-assisted operator clarification at Step~1.
Upon success, the CAMARA \ac{API} returns a Session ID and the confirmed \ac{QoS} parameters (\textbf{Step~6}).
This result is relayed back to the \ac{MCP} Client (\textbf{Step~7}), which feeds the structured output back into the \ac{LLM} context (\textbf{Step~8}).
Additionally, the \ac{LLM} synthesizes the final configuration payload in \ac{JSON} format, identifying the slice type (\textbf{Step~9}).
Finally, the \ac{MCP} Client transmits this structured intent to the ORION rApp within the \ac{Non-RT} \ac{RIC} for policy generation (\textbf{Step~10}).

\begin{figure*}[!h]
\centering
\includegraphics[width=0.8\linewidth]{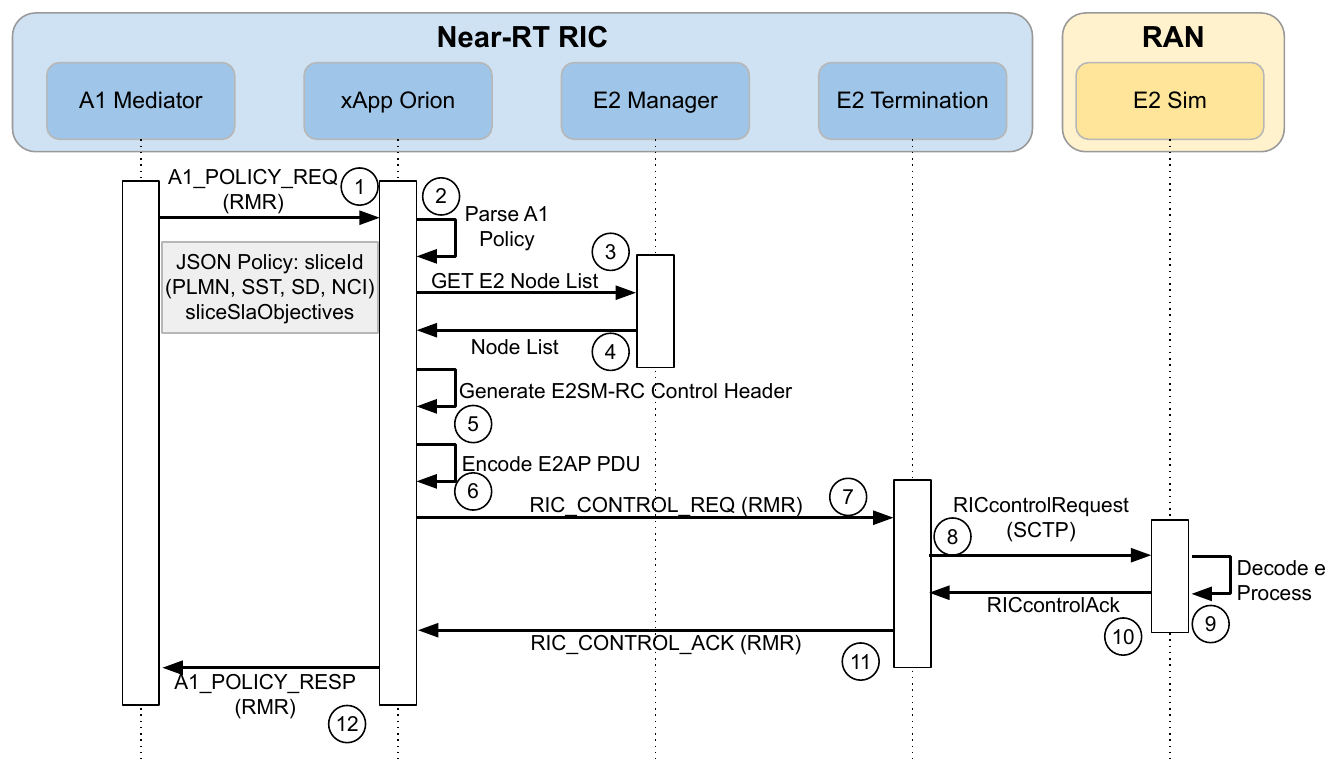}
\caption{Sequence diagram of the control loop.}
\label{fig:control-loop}
\end{figure*}

\subsection{ORION xApp}\label{sec:impl-xapp}
The ORION xApp is a C++ based application deployed on the \ac{Near-RT} \ac{RIC} platform.
Acting as the \ac{A1} Interface termination point, it receives \ac{JSON}-based policies from the \ac{Non-RT} \ac{RIC} via the \ac{A1} Mediator.
Upon receiving a policy, the ORION xApp is responsible for dynamically translating \ac{SLA} objectives into concrete resource allocations.
The ORION xApp retrieves the target \ac{E2} Node configuration, including bandwidth, numerology, and \ac{MIMO} layers, to estimate cell capacity.
The \ac{3GPP} throughput formulation depends on the modulation and coding scheme, as defined in TS~38.306~\cite{3GPP2024TS38306}.
Our prototype uses a proportional heuristic for resource allocation.
It maps requested throughput to \ac{PRB} percentage using the following linear approximation:
\begin{equation}
    \text{PRB}_{\%} = \left\lceil \frac{\text{Requested Throughput}}{\text{Max Cell Capacity}} \times 100 \right\rceil
\end{equation}
The xApp encodes the computed percentage into \ac{E2SM}-RC structures using \ac{ASN.1} \ac{PER} for \ac{E2AP} transmission.
The prototype applies these \ac{PRB} quotas to emulated scheduler partitions in \texttt{e2sim-rc}.

\subsection[E2Sim (e2sim-rc)]{E2Sim (\texttt{e2sim-rc})}\label{sec:impl-e2sim}
\texttt{e2sim-rc}~\cite{Huff2024E2SimRC} extends \ac{OSC} \ac{E2Sim} (\texttt{sim/e2-interface})~\cite{OSC2023E2Sim} for emulated \ac{gNB} slice control.\footnote{Source and deployment manifests: \url{https://github.com/Centro-Innova/e2sim-rc/tree/orion}.
ORION artifacts and sub-repositories: \url{https://github.com/Centro-Innova/analysis-orion}.}
It exchanges \ac{ASN.1} \ac{PER}-encoded \ac{E2AP} messages over \ac{SCTP} and advertises \ac{E2SM}-RC Control Style~2, Action ID~6 during \ac{E2} Setup with \ac{E2Mgr}.
\texttt{RICcontrolRequest} messages identify the slice and \ac{PRB} quota for emulated scheduler partitioning.
Allocations exceeding aggregate cell capacity return \texttt{RICcontrolFailure}, while valid allocations return \texttt{RICcontrolAcknowledge}.
\ac{E2SM}-KPM reports expose utilization and delay for the monitoring architecture.

\subsection{Near-RT RIC Workflow}\label{sec:impl-workflow-nearrt}

Figure~\ref{fig:control-loop} illustrates the interaction between the ORION xApp and the \ac{RAN} in the sequence diagram.
The process follows a policy deployment transaction triggered by the arrival of a new \ac{A1} policy.
The cycle begins when the \ac{A1} Mediator pushes a policy to the ORION xApp via the \ac{RMR} interface (\textbf{Step~1}).
The ORION xApp parses the \ac{JSON} payload to extract the target slice identifier and \ac{SLA} objectives (\textbf{Step~2}).
To determine enforcement targets, the ORION xApp queries the \ac{E2Mgr} (\textbf{Step~3}), which returns a list of connected \ac{E2} Nodes and their inventory (\textbf{Step~4}).
The ORION xApp's translation engine converts the parsed \ac{SLA} parameters into an \ac{E2SM}-RC Control Header and Control Message, specifying the Control Style (Type 2), Action ID (6), and \ac{RAN} Parameters for \ac{PRB} Policy Ratios (\textbf{Step~5}).
These structures are encapsulated into an \ac{E2AP} \texttt{RICcontrolRequest} \ac{PDU} and encoded into binary format (\textbf{Step~6}).

The encoded message is sent to the \ac{E2} Termination via \ac{RMR} (\textbf{Step~7}), which forwards it to the specific \ac{E2} Node over \ac{SCTP} (\textbf{Step~8}).
Upon receipt, the \texttt{e2sim-rc} decodes the payload and processes the request (\textbf{Step~9}), validating that the new allocation respects the node's capacity before updating the scheduler.
If successful, the simulator generates a \texttt{RICcontrolAcknowledge} message sent back to the \ac{E2} Termination (\textbf{Step~10}).
This acknowledgment is routed to the ORION xApp (\textbf{Step~11}) to confirm the configuration.
Finally, the ORION xApp reports operation status to the \ac{Non-RT} \ac{RIC} via an \ac{A1} Policy Response (\textbf{Step~12}), completing the deployment transaction.
\ac{E2SM}-KPM telemetry from \texttt{e2sim-rc} supports the monitoring path to the ORION xApp.
\ac{PRB} adjustment and \ac{A1} escalation on \ac{SLA} violations are architectural responses.

\section{Evaluation}\label{sec:evaluation}

This section evaluates the ORION framework in automating intent-driven network orchestration.\footnote{Analysis scripts, datasets, and figures are publicly available at: \url{https://github.com/Centro-Innova/analysis-orion}}
Following the experimental setup outlined in Section~\ref{subsec:experimental_setup}, our evaluation is structured around three primary dimensions.
First, we conduct \ac{LLM} benchmarking and comparative analysis (Section~\ref{sec:llm-analysis}), quantifying token consumption, policy creation success, cost-performance trade-offs, and slice-type classification accuracy.
Second, we perform resource consumption analysis (Section~\ref{subsec:resource_analysis}), measuring the computational overhead imposed by the orchestration architecture.
Third, we measure control-path processing latency (Section~\ref{subsec:latency_analysis}) through the emulated \ac{E2} node.

\subsection{Experimental Setup}\label{sec:eval-setup}
\label{subsec:experimental_setup}

To validate translation fidelity and orchestration performance, we use a curated evaluation dataset comprising 100 natural-language intents.
The archived dataset contains 34 \ac{eMBB}, 33 \ac{URLLC}, and 33 \ac{mMTC} intents, representing three \ac{5G} service families~\cite{ITU2015IMT2020}:
\begin{itemize}
    \item \textbf{\ac{eMBB} (34 samples):} High-throughput scenarios such as video journalism, cloud gaming, and 4K media streaming, prioritizing downstream rates (e.g., 100--500\,Mbps) and moderate latency.
    \item \textbf{\ac{URLLC} (33 samples):} Critical, low-latency applications, including autonomous robotics, remote surgery, and industrial automation, targeting strict packet delay budgets (1--7\,ms) and high availability.
    \item \textbf{\ac{mMTC} (33 samples):} Massive connectivity use cases, such as smart city sensors, environmental monitoring, and utility metering.
\end{itemize}

Dataset curation followed three stages.
(i) Seeds came from Dandoush et al.~\cite{dandoush2024largelanguagemodelsmeet}, ITU-R M.2083~\cite{ITU2015IMT2020}, and \ac{3GPP} TR~38.913~\cite{3GPP2018Scenarios}.
(ii) GPT-5 expanded them across linguistic structures, operational domains, and \ac{SLA} constraints.
(iii) Telecommunications domain experts audited every intent for ambiguity and feasibility.
The public archive provides intent identifiers, natural-language requests, reference slice-family labels, and tool-evaluation logs.\footnote{Publicly available at: \url{https://github.com/Centro-Innova/ORION/blob/main/src/test/intent_dataset.csv}}
Reference labels support deterministic slice-type classification checks, while the schema-aware judge evaluates generated tool arguments (Section~\ref{subsec:deepeval}).

\begin{table*}[!h]
\centering
\caption{Composition of the 100-intent benchmark with verbatim examples.}
\label{tab:dataset_composition}
\renewcommand{\arraystretch}{1.15}
\resizebox{\textwidth}{!}{%
\begin{tabular}{lc >{\raggedright\arraybackslash}p{4.2cm} >{\raggedright\arraybackslash}p{10.2cm}}
\hline
\cellcolor{DarkBlue!80}\textbf{Slice Family} &
\cellcolor{DarkBlue!80}\textbf{Samples} &
\cellcolor{DarkBlue!80}\textbf{Explicit Example Attributes} &
\cellcolor{DarkBlue!80}\textbf{Representative Natural-Language Intent Example} \\
\hline
\rowcolor{white}
\ac{mMTC} &
33 (33\%) &
3,000 devices; uplink 1\,Mbps per device &
\textit{``Create a slice for environmental sensors monitoring air quality across a city district, supporting up to 3000 devices with uplink rates of 1 Mbps per device.''} \\
\rowcolor{LightBlue!100}
\ac{eMBB} &
34 (34\%) &
40 devices; downstream 120\,Mbps per device; packet delay budget 15\,ms &
\textit{``Create a slice to support video journalists transmitting live footage from a downtown area, allowing up to 40 devices with downstream rates of 120 Mbps per device and a packet delay budget of 15 ms to keep streams smooth.''} \\
\rowcolor{white}
\ac{URLLC} &
33 (33\%) &
Packet delay budget 3\,ms; uplink 20\,Mbps per device &
\textit{``Create a slice to control autonomous delivery robots in an industrial zone, requiring packet delay budget of 3 ms and uplink rates of 20 Mbps per device.''} \\
\hline
\end{tabular}%
}
\end{table*}

Intents comprise imperative (45\%), declarative (35\%), and conversational (20\%) forms, with lengths of 12--58 tokens ($28.4 \pm 9.6$).
Terminology ranges from formal \ac{3GPP} parameters to colloquial application descriptions (Table~\ref{tab:dataset_composition}).
60\% of the supply is complete numerical parameters, while 40\% omits non-mandatory attributes.
Partial intents test null preservation and downstream \ac{3GPP} baseline-profile binding in the ORION rApp.

100 intents concern slice provisioning and reservation across \ac{mMTC}, \ac{eMBB}, and \ac{URLLC}.
The benchmark covers Creation and Activation within the architectural lifecycle informed by \ac{3GPP} TS~28.312~\cite{3GPP2025Intent}.

Each entry in the dataset pairs a realistic, unstructured operator prompt with its reference slice-family label.
This dataset serves as the input workload for the \ac{SMO} layer to quantify the \ac{LLM}'s ability to correctly extract and map intents to the \ac{CAMARA}-compliant \textit{NetworkSliceBooking} schema.

The experimental testbed was hosted on a Kubernetes cluster deployed on a VMware-virtualized host running Ubuntu 24.04.2 LTS.
The virtualized infrastructure was provisioned with 12\,vCPUs (Intel Xeon X5660 @ 2.80\,GHz) and 23\,GiB of RAM, representing resource-constrained legacy edge infrastructure.
This environment hosted the full ORION stack, including the \ac{SMO} framework, \ac{Non-RT} and \ac{Near-RT} \ac{RIC} instances, and the RAN simulation components.

\subsection{LLM Benchmarking and Comparative Analysis}\label{sec:eval-llm-benchmarking}
\label{sec:llm-analysis}

We benchmarked eight \acp{LLM} using the same ORION workflow and 100-intent workload.
The benchmark covers Claude Opus~4.5, Claude Sonnet~4.5, Gemini~3 Flash, Gemini~3 Pro, GPT-5, GPT-5 Nano, Qwen3-235B-A22B-Instruct-2507, and Qwen3-30B-A3B-Instruct-2507.
Anthropic, Google, and OpenAI provide the Claude, Gemini, and GPT models, respectively.
DeepInfra and SiliconFlow host Qwen3-235B (235B total parameters, 22B active per token)~\cite{Qwen2025Qwen3TwoThirtyFiveB} and Qwen3-30B (30.5B total, 3.3B active)~\cite{Qwen2025Qwen3ThirtyB}, respectively, covering frontier-scale and smaller open-weight checkpoints.
Each model processed identical \ac{O-RAN} network slice intents spanning \ac{eMBB}, \ac{URLLC}, and \ac{mMTC} use cases, with the \ac{SMO} orchestrating tool calls to translate natural language requests into \ac{A1} policies.
Table~\ref{tab:llm-performance} summarizes the key performance metrics.
$N$ denotes the number of intents represented in the parsed usage records: 100/100 for Claude, GPT, and Qwen models, 77/100 for Gemini 3 Flash, and 78/100 for Gemini 3 Pro.

Each model run processes the 100-intent dataset sequentially through the \texttt{/intent} endpoint with a 60-second pause between requests.
The instruction prompt supplies the same eight field-population rules, and all runs use identical \ac{MCP} tool schemas.
The \ac{MCP} Server validates argument structure with the tool's Pydantic model before forwarding the booking request.
The workflow translates each intent into a booking tool call, classifies the slice type, and submits the resulting specification to the ORION rApp for policy creation.

Translation and subsequent classification token usage are summed per observed intent.
Reported means and error bars reflect per-intent aggregates and sample standard deviations across the observed intents.
Generation costs exclude judge evaluation and use the inference stages recorded for each intent.
Malformed arguments, missing tool calls, and absent predictions are retained as recorded failure outcomes.

    \begin{table*}[!h]
    \centering
    \caption{LLM Performance Comparison for Intent Processing}
    \label{tab:llm-performance}
    \renewcommand{\arraystretch}{1.15}
    \resizebox{.8\textwidth}{!}{%
    \begin{tabular}{lrrrcccc}
    \hline
    \cellcolor{DarkBlue!80}\textbf{Model} & \cellcolor{DarkBlue!80}\textbf{N} & \cellcolor{DarkBlue!80}\textbf{Input Tokens} & \cellcolor{DarkBlue!80}\textbf{Output Tokens} & \cellcolor{DarkBlue!80}\textbf{Reasoning (\%)} & \cellcolor{DarkBlue!80}\textbf{Success (\%)} & \cellcolor{DarkBlue!80}\textbf{Cost (cents/intent)} \\
    \hline
    \rowcolor{white}
    Claude Opus 4.5   & 100 & 18,766 $\pm$ 9,616  & 204 $\pm$ 115    & N/A  & 100.0 & 29.68 \\
    \rowcolor{LightBlue!100}
    Claude Sonnet 4.5 & 100 & 15,270 $\pm$ 9,679  & 154 $\pm$ 16     & N/A  & 17.0  & 4.81  \\
    \rowcolor{white}
    Gemini 3 Flash    & 77  & 15,090 $\pm$ 8,308  & 2,418 $\pm$ 2,028 & 23.1 & 85.7  & 0.19  \\
    \rowcolor{LightBlue!100}
    Gemini 3 Pro      & 78  & 15,226 $\pm$ 8,173  & 2,394 $\pm$ 2,787 & 21.0 & 100.0 & 3.10  \\
    \rowcolor{white}
    GPT-5 Nano        & 100 & 26,349 $\pm$ 14,797 & 1,512 $\pm$ 381   & 4.9  & 97.0  & 0.49  \\
    \rowcolor{LightBlue!100}
    GPT-5             & 100 & 27,306 $\pm$ 15,138 & 775 $\pm$ 226     & 2.4  & 100.0 & 7.60  \\
    \rowcolor{white}
    Qwen3 235B A22B   & 100 & 21,546 $\pm$ 10,754 & 84 $\pm$ 9       & 0.0 & 8.0 & 0.1986 \\
    \rowcolor{LightBlue!100}
    Qwen3 30B A3B     & 100 & 29,568 $\pm$ 16,023 & 261 $\pm$ 27     & 0.0 & 3.0 & 0.2739 \\
    \hline
    \end{tabular}%
    }
    \end{table*}

\subsubsection{Token Consumption Patterns}\label{sec:eval-tokens}
\begin{figure}[!h]
\centering
\includegraphics[width=1\columnwidth]{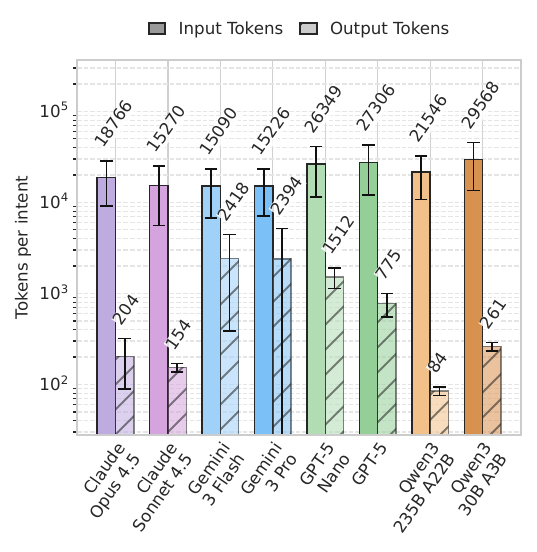}
\caption{Token consumption per intent across LLM providers (log scale). Solid bars represent input tokens, while hatched bars represent output tokens. Error bars indicate sample standard deviation over the observed $N$ intents in Table~\ref{tab:llm-performance}.}
\label{fig:llm-tokens}
\end{figure}

Input token consumption varies significantly across models, ranging from 15,090 tokens for Gemini 3 Flash to 29,568 for Qwen3-30B (Fig.~\ref{fig:llm-tokens}).
This variance reflects differences in tokenization strategies, schema representations, and context window utilization.
OpenAI and Qwen models exhibit higher input consumption, with Qwen3-235B recording $21{,}546 \pm 10{,}754$ and Qwen3-30B recording $29{,}568 \pm 16{,}023$ input tokens per attempted intent across all 100 benchmark attempts.
Output tokens show even greater variance: Gemini models produce 2,394--2,418 output tokens on average due to verbose chain-of-thought reasoning, compared to 154--1,512 for Anthropic and OpenAI models.
In contrast, Qwen3-235B and Qwen3-30B produce only $84 \pm 9$ and $261 \pm 27$ output tokens per attempted intent, respectively.

\subsubsection{Reasoning Token Allocation}\label{sec:eval-reasoning}
\begin{figure}[!h]
\centering
\includegraphics[width=1\columnwidth]{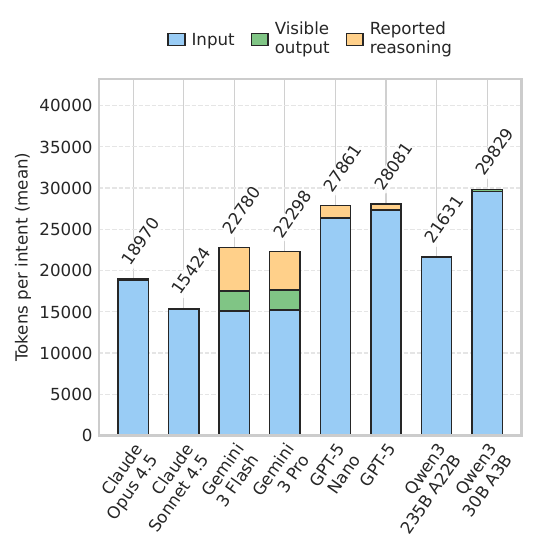}
\caption{Input, visible output, and reported reasoning tokens per observed intent.
Provider accounting conventions are described in the text.}
\label{fig:llm-breakdown}
\end{figure}

The reasoning ratio is computed as $100R/(I+O+R)$, where $I$, $O$, and $R$ denote reported input, output, and reasoning counters, respectively (Fig.~\ref{fig:llm-breakdown}).
Models exposing reasoning tokens exhibit distinct deliberation strategies: Gemini models allocate 21--23\% of their token budget to explicit reasoning, compared to 2.4--4.9\% for GPT-5 variants.
This disparity indicates deliberate chain-of-thought processing in Gemini, whereas GPT-5 models rely on more direct inference patterns.

For OpenAI and Qwen, visible output subtracts reported reasoning from inclusive output counters to avoid double counting (Fig.~\ref{fig:llm-breakdown}).
The reasoning ratio remains an accounting convention rather than a partition of disjoint tokens.
Anthropic models do not expose reasoning counters in their API responses, precluding direct comparison of internal deliberation overhead.
Hosted Qwen responses report zero reasoning tokens; this recorded provider counter reflects API reporting conventions rather than an independently established absence of internal deliberation.

\subsubsection{Policy Creation Success}\label{sec:eval-success}
\begin{figure}[!h]
\centering
\includegraphics[width=0.95\columnwidth]{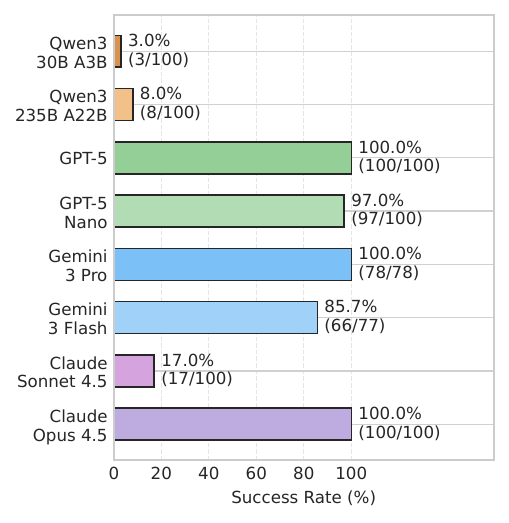}
\caption{Policy creation success over the observed $N$ intents in Table~\ref{tab:llm-performance}.
A successful rApp response defines policy creation.}
\label{fig:llm-success}
\end{figure}

Policy creation success is determined by the recorded rApp policy response, independently of reference labels and judge scores (Fig.~\ref{fig:llm-success}).
We evaluate success percentages conditional on the observed intents in Table~\ref{tab:llm-performance}.
Four models achieved $\geq$ 97\% success: Claude Opus 4.5 (100\%), Gemini 3 Pro (78/78 observed, 100\%; 78\% of the full workload), GPT-5 (100\%), and GPT-5 Nano (97\%).
Gemini 3 Flash achieved 85.7\% (66/77 observed; 66\% overall) due to occasional malformed tool invocations.
Claude Sonnet 4.5 achieves 17\% success, while Qwen3-235B and Qwen3-30B create 8/100 and 3/100 policies, respectively.

The remaining 92 and 97 Qwen outcomes contain text without recorded \ac{MCP} calls or policies.
HTTP 200 responses do not establish policy creation.
Paralleling Claude Sonnet 4.5's tool invocation failures, these outcomes demonstrate that fluent language understanding does not guarantee reliable agentic tool orchestration in the \ac{SMO}.

\subsubsection{Cost-Performance Tradeoff}\label{sec:eval-cost}
\begin{figure}[!h]
\centering
\includegraphics[width=0.95\columnwidth]{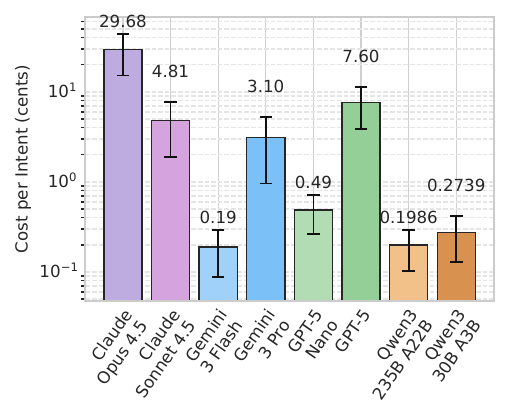}
\caption{Generation API cost per observed intent (cents, log scale).
Error bars indicate sample standard deviation.}
\label{fig:llm-cost}
\end{figure}

Generation API costs vary across models (Fig.~\ref{fig:llm-cost}).
Mean costs range from 0.19\,cents per intent for Gemini~3 Flash to 29.68\,cents for Claude Opus~4.5, a 156$\times$ difference.
Qwen3-235B and Qwen3-30B record mean costs of $0.199 \pm 0.097$ and $0.274 \pm 0.144$\,cents per observed intent, respectively.

Generation API cost alone does not predict recorded policy-creation success.
For an \ac{O-RAN} system processing 10,000 daily intents, annual API charges could range from \$6,935 (Gemini 3 Flash) to \$1,083,320 (Claude Opus 4.5).
Operational viability requires weighting raw token tariffs against recorded policy creation.

\subsubsection{Slice-Type Classification Accuracy}\label{sec:eval-classification-accuracy}
\label{subsec:classification-accuracy}
\begin{figure}[!h]
\centering
\includegraphics[width=0.95\columnwidth]{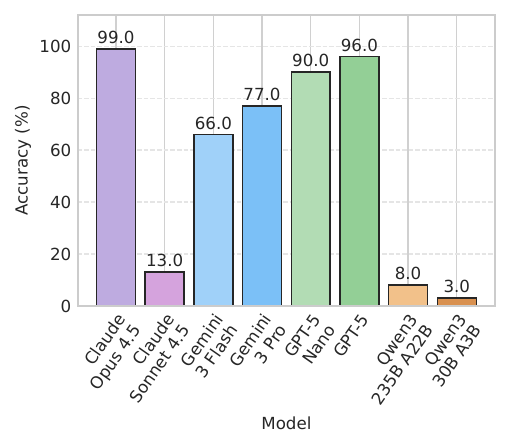}
\caption{Slice-type classification accuracy across the 100-intent benchmark, including missing predictions.}
\label{fig:slice-accuracy}
\end{figure}

Beyond policy creation success, we evaluate whether generated configurations match the \ac{3GPP} slice family (\ac{eMBB}, \ac{URLLC}, or \ac{mMTC}) implied by the intent.
Classification accuracy is evaluated deterministically against the dataset reference label across all benchmark intents, including missing predictions (Fig.~\ref{fig:slice-accuracy}).
Three performance tiers emerge among models that reliably produce predictions: Claude Opus 4.5 leads with 99\% accuracy, followed by GPT-5 at 96\% (100/100 coverage) and GPT-5 Nano at 90\% (97/100 coverage).
Gemini 3 Pro achieves 77\% (78/100 coverage), while Gemini 3 Flash achieves 66\% overall (66/100 coverage) and 100\% precision on intents where it produces predictions (66/66).

In the lowest tier, overall reference-label agreement is constrained by tool orchestration failures rather than semantic misclassification.
Claude Sonnet 4.5 yields 13\% accuracy, reflecting a 17/100 prediction rate due to tool invocation omissions.
Qwen3-235B and Qwen3-30B have matching overall agreement and prediction coverage of 8/100 and 3/100, respectively.
All predictions are correct (8/8 and 3/3) but cover only eMBB, with no URLLC or mMTC classifications because booking invocations are absent.

\subsubsection{MCP Tool-Use Quality Assessment}\label{sec:eval-mcp-quality}
\label{subsec:deepeval}
To evaluate the parameter fidelity of generated tool calls against the \textit{NetworkSliceBooking} schema, we employ DeepEval's \texttt{MCPUseMetric}~\cite{Ip2024DeepEval, Qin2024ToolLLM}.
Claude Opus 4.6 judges the responses across all eight models at temperature 0 with a 1,024-token limit.
Each test case supplies the system prompt rules and user intent, with the model's response evaluated against the ORION \ac{MCP} Server descriptor.
The metric assesses tool selection and parameter fidelity on a continuous 0--1 scale.

\begin{figure}[!h]
\centering
\includegraphics[width=0.95\columnwidth]{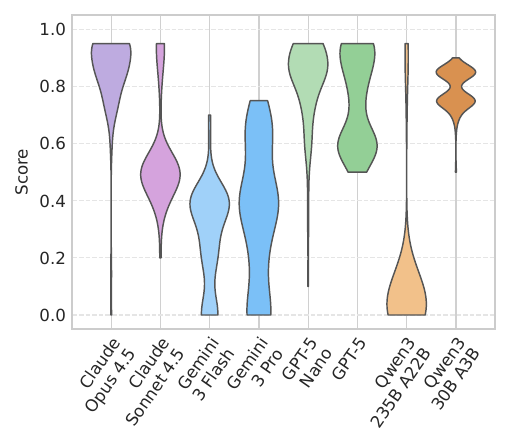}
\caption{DeepEval \texttt{MCPUseMetric} score distributions across evaluated LLM providers.
Violin plots show kernel density estimates.}
\label{fig:deepeval-dist}
\end{figure}

Figure~\ref{fig:deepeval-dist} presents the resulting score distributions across the eight models.
Claude Opus 4.5 and GPT-5 Nano form a high-performance cluster with mean scores of 0.85 and 0.82, respectively, followed by GPT-5 at 0.75.
Gemini models occupy a lower tier (Gemini 3 Pro at 0.36, Gemini 3 Flash at 0.28), while Claude Sonnet 4.5 averages 0.55.
Qwen3-235B and Qwen3-30B record mean judge scores of $0.11 \pm 0.24$ and $0.79 \pm 0.06$, respectively.

Notably, Qwen3-30B achieves a high mean judge score because its text-only outputs contain plausible parameter values despite omitting actual \ac{MCP} tool invocations.
Failure analysis across the remaining models reveals three primary deduction causes.
Gemini models generate duplicate session calls in truncated formats, Claude Sonnet 4.5 refuses optional nullable fields, and GPT-5 places generic delay budgets ambiguously across directional fields.
These patterns show that response-level judge scores complement, rather than replace, verified rApp policy-creation outcomes.

\subsection{Resource Consumption Analysis}\label{sec:eval-resources}
\label{subsec:resource_analysis}

\begin{figure}[!h]
    \centering
    \includegraphics[width=\linewidth]{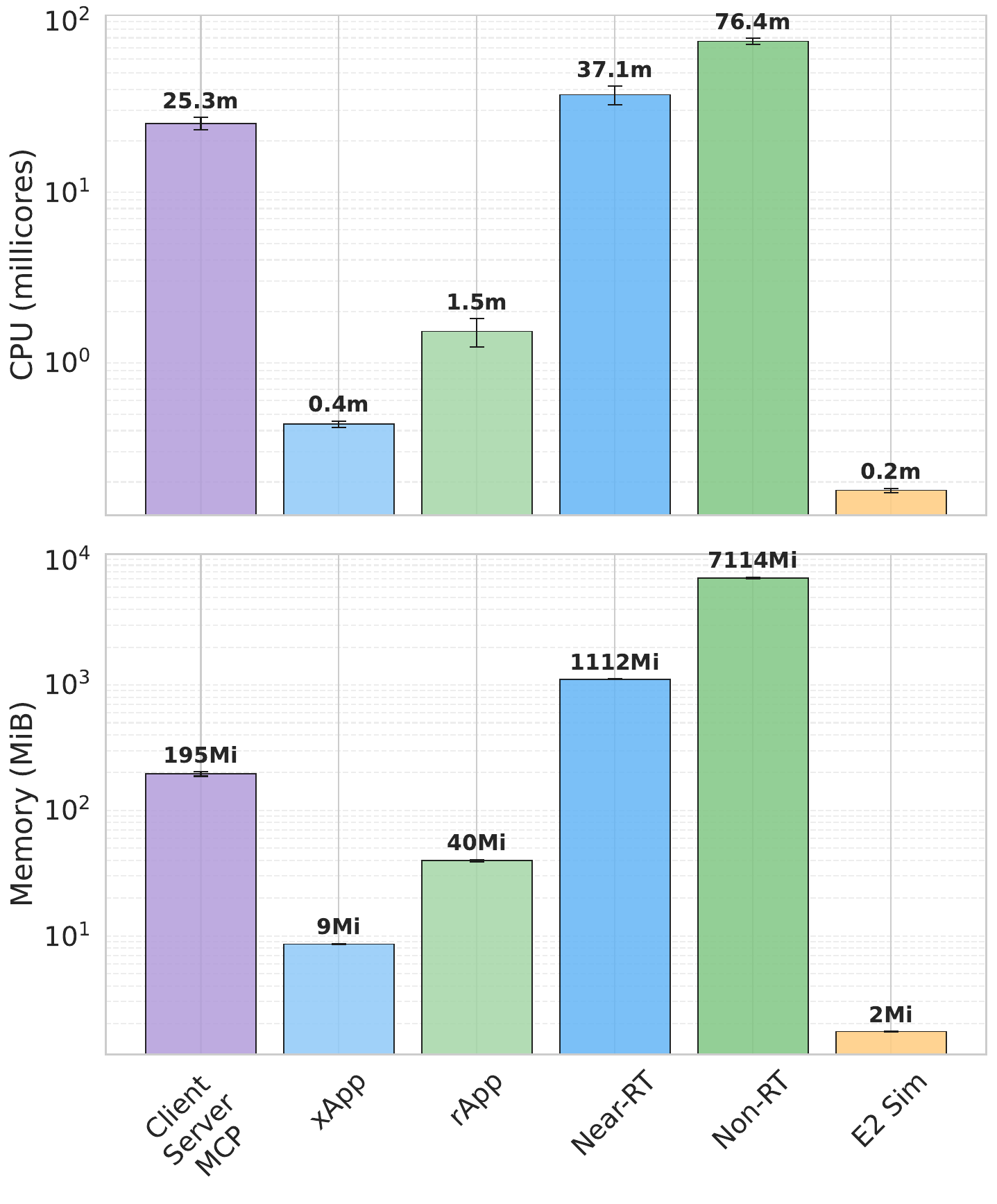} 
    \caption{CPU (millicores) and Memory (MiB) usage by namespace (Mean $\pm$ Std Dev).}
    \label{fig:ns_resource_usage}
\end{figure}

We assessed the computational overhead of the proposed intent orchestration framework using the Prometheus monitoring stack to scrape CPU and memory metrics from the target Kubernetes namespaces.
The evaluation captured the steady-state performance of the \ac{O-RAN} platforms (\ac{Non-RT} \ac{RIC}, \ac{Near-RT} \ac{RIC}) and the developed \ac{MCP} Server/Client, alongside the specific intent-handling components (ORION rApp, ORION xApp) and the \ac{E2Sim}.
Figure~\ref{fig:ns_resource_usage} shows resource use by the infrastructure platforms and orchestration components.

The quantitative results confirm the efficiency of our approach.
The \ac{Non-RT} \ac{RIC} remains the most resource-intensive component, consuming an average of $76.4 \pm 3.4$\,m CPU and $\approx 7.1$\,GiB of memory due to its comprehensive microservice suite.
The \ac{Near-RT} \ac{RIC} consumes $37.1 \pm 4.6$\,m CPU and $1.1$\,GiB memory.
The ORION rApp and ORION xApp consume $1.5 \pm 0.3$\,m and $0.4 \pm 0.0$\,m CPU, respectively, with memory footprints under $40$\,MiB.
The \ac{MCP} Client/Server and \ac{E2Sim} account for the remaining $\approx 25.6$\,m CPU overhead.
The measured total system-wide draw is approximately $141$\,millicores and $8.5$\,GiB RAM on this testbed.

\subsection{End-to-End Latency Analysis}\label{subsec:latency_analysis}\label{sec:eval-latency}

\begin{figure}[!h]
    \centering
    \includegraphics[width=1.\linewidth]{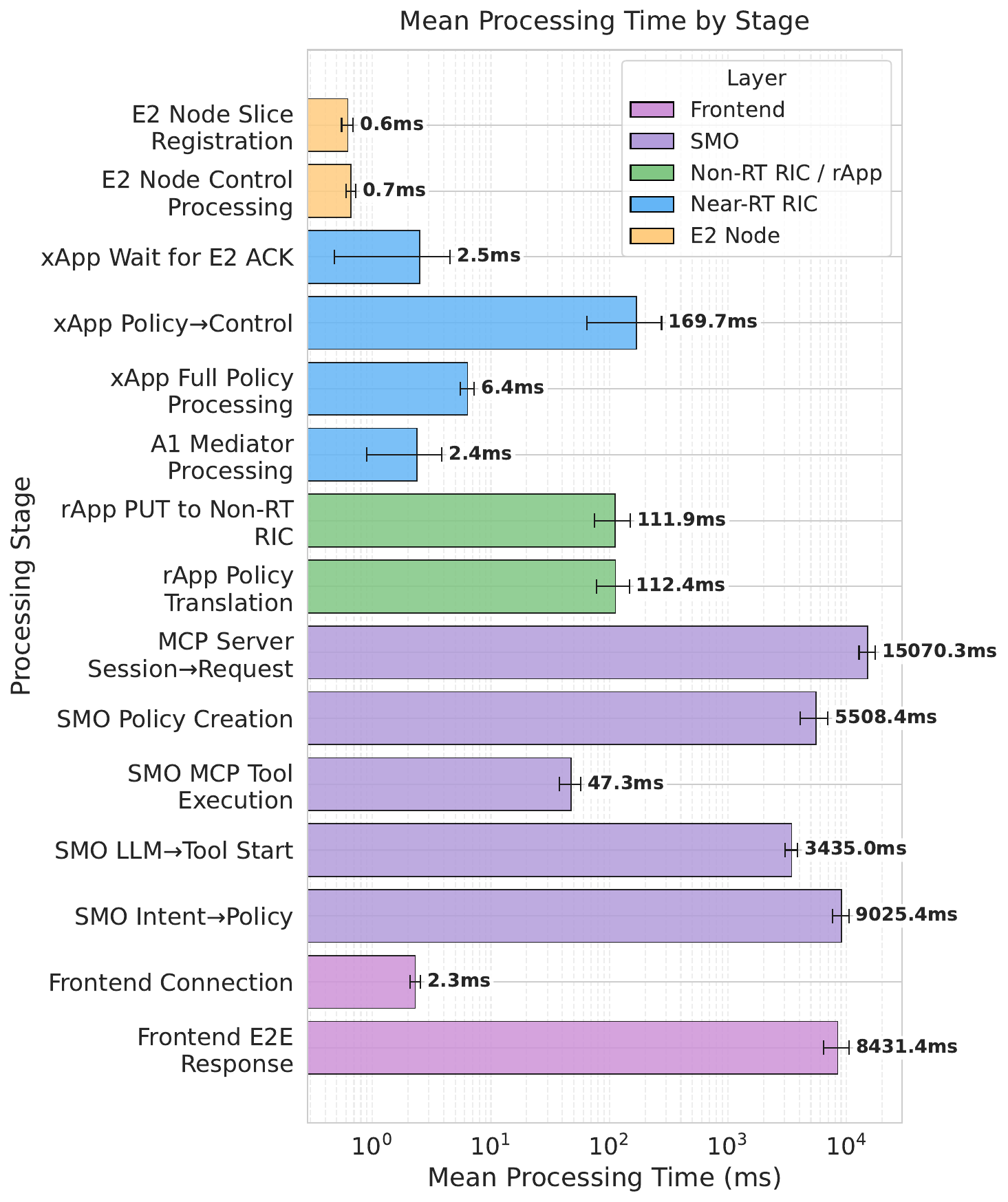}
    \caption{Mean processing time by stage across SMO, Near-RT RIC, and E2 Node.
    Error bars indicate standard deviation.}
    \label{fig:proc_time}
\end{figure}

Latency measurements follow GPT-5 intent processing from front-end submission through policy generation and the emulated \ac{E2} control path (Fig.~\ref{fig:proc_time}).
Mean processing times and standard deviations describe the recorded stages in this prototype.
The \ac{SMO} stages, including \textit{\ac{SMO} Intent$\to$Policy} and \textit{\ac{SMO} \ac{MCP} Tool Execution}, account for most of the measured delay.
The \textit{xApp Full Policy Processing} and \ac{A1} Mediator stages contribute smaller delays.
The \textit{xApp Policy$\to$Control} and \textit{\ac{E2} Node Control Processing} stages operate in the reported sub-second to millisecond range.
These results characterize control-message processing, including \texttt{e2sim-rc} handling of emulated \ac{PRB} allocations.

\subsection{Discussion: Applicability and Limitations}\label{sec:eval-discussion}
Translating unconstrained, colloquial, and partially specified operator intents into schema-validated \ac{CAMARA} \textit{NetworkSliceBooking} payloads is a non-trivial task that justifies capable \acp{LLM} in dynamic, multi-tenant \ac{O-RAN} environments.
While deterministic, rule-based orchestration suffices for static provisioning, on-the-fly slice creation requires semantic slot extraction, unit normalization, and strict null preservation across the 40\% of intents that omit optional attributes.
The model must also invoke structured \ac{MCP} tools and classify the \ac{3GPP} slice family (\ac{eMBB}, \ac{URLLC}, or \ac{mMTC}) to guide xApp scheduler selection.
Our benchmark confirms this complexity empirically: four frontier cloud configurations achieved 97--100\% policy creation, whereas Claude Sonnet~4.5, Qwen3-235B, and Qwen3-30B succeeded on only 3--17\% of intents due to omitted \ac{MCP} calls or schema violations.

Sending operator intents to third-party cloud \acp{LLM} introduces confidentiality, data-residency, and external availability risks.
ORION bounds controller exposure by keeping backend credentials, internal \ac{SMO}/\ac{RIC} endpoints, Pydantic schema checks, and deterministic \ac{A1}/\ac{E2} generation within the operator-hosted \ac{MCP} Server, rApp, and xApp.
Because the \ac{MCP} Client is model-agnostic, operators can substitute on-premises \ac{LLM} endpoints without modifying the \ac{SMO}, rApp, or xApp.
In this study, we evaluated open-weight Qwen3-30B and Qwen3-235B through a hosted \ac{API} because these checkpoints achieved only 3--8\% policy creation on our strict tool-calling workload.
Moreover, our legacy testbed (12\,vCPUs, 23\,GiB RAM) lacked the local GPU capacity required to self-host larger open-weight checkpoints.
Future work will therefore evaluate on-premises open-weight deployments, domain-specific fine-tuning for \ac{MCP} tool invocation, local inference latency, and hardware sizing.

%
%
\section{Conclusion}\label{sec:conclusion}
We presented ORION, an intent orchestration framework that maps operator objectives to \ac{A1} policies through the \ac{CAMARA} \textit{NetworkSliceBooking} schema.
The prototype implements slice provisioning and activation with \ac{E2SM}-RC control of \ac{PRB} quotas.
The evaluation covers eight model configurations, including two open-weight models, across a 100-intent workload.

Claude Opus 4.5 and GPT-5 achieved 100\% policy-creation success, while GPT-5 Nano achieved 97\% at 0.49\,cents per intent.
Their slice-type classification accuracies were 99\%, 96\%, and 90\%, respectively, complementing the \ac{MCP} tool-quality evaluations.
Hosted Qwen3-235B and Qwen3-30B created 8/100 and 3/100 policies, respectively.
Resource measurements show small rApp and xApp overhead, while \ac{LLM} processing at the \ac{SMO} dominates the measured GPT-5 control-path latency.

Future work will extend the evaluation to local \ac{LLM} deployments, assessing trade-offs between inference latency, privacy implications, and reasoning capability compared to cloud-hosted models.
We also plan to explore the \ac{AI-and-RAN} pillar of the \ac{AI-RAN} vision, investigating how the computational resources used to run ORION's \ac{LLM} instances can be dynamically shared with edge user workloads.
Finally, while our prototype evaluates slice provisioning, activation, and \ac{E2SM}-RC \ac{PRB} control, future work will evaluate interactive clarification dialogues, closed-loop runtime adaptation and rollback from real-time \ac{RAN} telemetry, and cross-operator federation.

\section*{Acknowledgment}
The research leading to this article received support from the Program OpenRAN@Brasil. This work was also partially funded by the project XGM-AFCCT-2026-5-1-1 supported by xGMobile - EMBRAPII - Inatel Competence Center on 5G and 6G Networks, with financial resources from the PPI IoT/Manufatura 4.0 / the MCTI grant number 052/2023, signed with EMBRAPII; the National Council for Scientific and Technological Development (CNPq); and the MCTIC/CGI.br/São Paulo Research Foundation (FAPESP) through project Programmability, ORchestration and VIRtualization of 5G Networks (PORVIR-5G) (Grants No.\ 2020/05182-3 and 2025/01970-0).

\ifCLASSOPTIONcaptionsoff
  \newpage
\fi



%
\bibliographystyle{IEEEtran}
\bibliography{bibtex/bib/IEEEabrv,bibtex/bib/refs}

\begin{IEEEbiography}[{\includegraphics[width=1in,height=1.25in,clip,keepaspectratio]{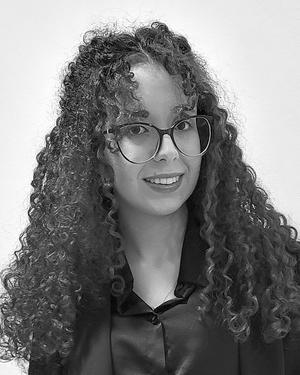}}]{Gabriela da Silva Machado}
received the B.S. degree in systems analysis and development and the M.Sc. degree in applied computing from the University of Vale do Rio dos Sinos (UNISINOS), Brazil.
Her research interests include Open RAN, intent-based networking, and AI-driven slice management.
\end{IEEEbiography}

\begin{IEEEbiography}[{\includegraphics[width=1in,height=1.25in,clip,keepaspectratio]{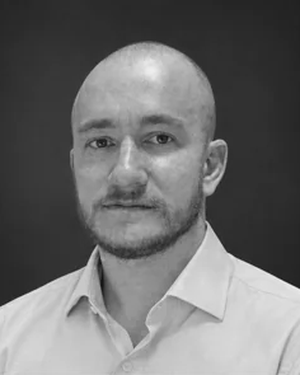}}]{Gustavo Z. Bruno}
received the Ph.D. degree from UNISINOS, Brazil, with a visiting research period at Virginia Tech, USA.
He is currently a Postdoctoral Researcher with INATEL, Brazil.
His research focuses on Open RAN, large language models, and intent-driven network orchestration.
\end{IEEEbiography}

\begin{IEEEbiography}[{\includegraphics[width=1in,height=1.25in,clip,keepaspectratio]{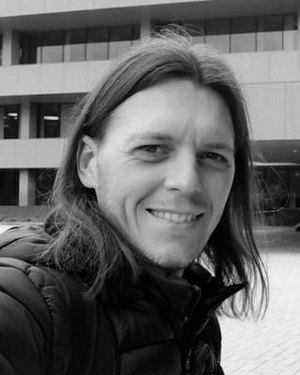}}]{Alexandre Huff}
received the Ph.D. degree in computer science from the Federal University of Paran{\'a} (UFPR), Brazil, in 2021.
He is an Associate Professor with the Federal University of Technology -- Paran{\'a} (UTFPR), Toledo, Brazil.
His research focuses on O-RAN architectures, E2 service models, and radio resource allocation.
\end{IEEEbiography}

\begin{IEEEbiography}[{\includegraphics[width=1in,height=1.25in,clip,keepaspectratio]{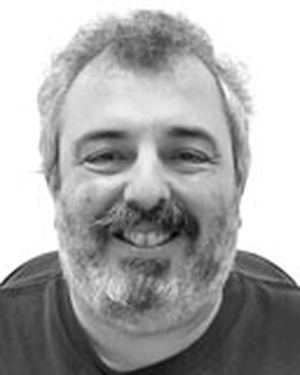}}]{Jos{\'e} Marcos C. Brito}
received the B.S. degree from INATEL and the M.Sc. and Ph.D. degrees in electrical engineering from UNICAMP, Brazil.
He is a Full Professor with the National Telecommunications Institute (INATEL), Santa Rita do Sapuca{\'i}, Brazil.
His research interests include wireless networks, 5G/6G architectures, and mobile communication systems.
\end{IEEEbiography}

\begin{IEEEbiography}[{\includegraphics[width=1in,height=1.25in,clip,keepaspectratio]{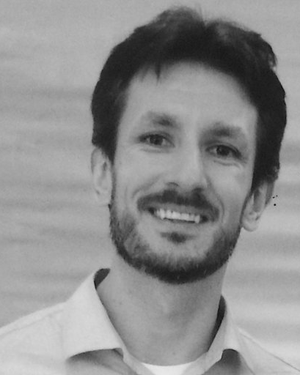}}]{Cristiano B. Both}
received the Ph.D. degree in computer science from the Federal University of Rio Grande do Sul (UFRGS), Brazil.
He is a Full Professor with the Applied Computing Graduate Program at the University of Vale do Rio dos Sinos (UNISINOS), Brazil.
His research focuses on softwarized networks, Open RAN orchestration, and cognitive mobile systems.
\end{IEEEbiography}




\end{document}